\documentclass[12pt]{article}     \usepackage{amsmath}
\usepackage{amsthm}               \usepackage{amsopn}
\usepackage{graphics}
\usepackage{latexsym}             \usepackage{amsfonts}
\usepackage{amssymb}
\usepackage{amsmath}
\usepackage{enumerate}
\usepackage{array}
\usepackage{graphicx}
\usepackage{amsopn}
\usepackage[toc,page]{appendix}
\usepackage[T1]{fontenc}
\usepackage[utf8]{inputenc}
\DeclareUnicodeCharacter{2003}{-}
\usepackage{authblk}
\usepackage{footnote}

\swapnumbers
\theoremstyle{plain}

\theoremstyle{definition}

\def\qed{\hfill\rule{1ex}{1ex}\\}

\title{Evaluating the Effectiveness of Health Awareness Events
by Google Search Frequency}
\newcommand*\samethanks[1][\value{footnote}]{\footnotemark[#1]}

\author[1]{Zheng Hao\thanks{These authors contributed equally to this work.}}
\author[2]{Miao Liu\samethanks}
\author[2]{Xijin Ge\thanks{xijin.ge@sdstate.edu}}

\affil[1]{State University of New York at Oswego}
\affil[2]{South Dakota State University}

\newcommand{\nc}[2]{ \newcommand{#1}{#2} }

\nc{\avint}{ {- \hspace{-3.5mm} \int} }  
\nc{\xpr}{x^{\prime}}

\nc{\R}{{\rm {I \! R}}}  
\nc{\N}{{\rm {I \! N}}}  

\newcommand{\newsec}[2]{ \section{#1} \label{sec-#2}  
                         \setcounter{equation}{0}     
                         \setcounter{theorem}{0} }    
\newcommand{\newsub}[2]{ \subsection{#1} \label{sub-#2} }

\nc{\Holder}{H\"{o}lder\ }

\nc{\ith}{ \ensuremath{\text{i}^{\text{th}}} }
\nc{\jth}{ \ensuremath{\text{j}^{\text{th}}} }
\nc{\kth}{ \ensuremath{\text{k}^{\text{th}}} }
\nc{\curl}{ \nabla \times }
\nc{\Div}{ \nabla \cdot }

\nc{\Ppl}{ \mathcal{M}^{+} }  \nc{\Pmn}{ \mathcal{M}^{-} }

\nc{\smiley}{ $\stackrel{\because}{\smile} \;$ }

\begin{document}
\numberwithin{equation}{section}
\maketitle

\begin{abstract} \noindent
Over two hundreds health awareness events take place in the United States in order to raise attention and educate the public about diseases. It would be informative and instructive for the organization to know the impact of these events, although such information could be difficult to measure. Here 46 events are selected and their data from 2004 to 2017 are downloaded from Google Trend(GT). We investigate whether the events effectively attract the public attention by increasing the search frequencies of certain keywords which we call queries. Three statistical methods including Transfer Function Noise modeling, Wilcoxon Rank Sum test, and Binomial inference are conducted on 46 GT data sets. Our study show that 10 health awareness events are effective with evidence of a significant increase in search frequencies in the event months, and 28 events are ineffective, with the rest being classified as unclear.\ 

\end{abstract}

\newsec{Introduction}{Intro}
\newsub{Background}{bg}
Chronic diseases (such as diabetes, cancer and heart diseases) cause $70\%$ of deaths in the United States every year, even though many of those diseases are preventable \cite{1}. The goal of holding health awareness events is to raise attention and educate the public about diseases. Take the National Breast Cancer awareness month as an example: the National Breast Cancer Foundation devotes efforts to educating women on early detection to reduce the risk of breast cancer, helping those diagnosed with breast cancer, as well as raising funds to support research. Companies join the National Breast Cancer Awareness Month, such as Estée Lauder Companies Inc. who releases exclusive Pink Ribbon products to help improve awareness of breast cancer and raise funds for medical research \cite{2}.

It is estimated that $97\%$ of the information flowing through two-way telecommunication were carried by the Internet by 2007 \cite{3}. The number of Internet users has increased enormously and surpasses 3 billion or about $46.1\%$ of the world population in 2014 \cite{4}. Google has led the U.S. core search market for the past decade \cite{5}, and millions of people worldwide use it to search for health topics every day \cite{6}\cite{7}. In particular it occupied three quarters of the search engine market in 2017. 
We want to determine if effective health awareness events are effective in raising public awareness of the health topic resulting in higher Google search frequencies. The results could benefit a variety of parties, for instance, the Department of Public Health and public interest groups could optimally rearrange resources allocation among events.

\newsub{Related Work}{LR}
Using Internet statistics to explain and predict quantities has been popular among researcher. Bollen et al.\cite{8} classified tweets into different moods to quantify the daily public mood and used it to predict stock market by using different models. The idea was based on the fact that people intentionally or unintentionally disclosed their thinking online by some means including social media such as Twitter, which might be a factor of stock price variation. What was interesting was that the authors used tweets which was not traditionally considered as an economic factor unlike some classical factors such as interest rates, GDP, and unemployment rates. 



Ginsberg et al.\cite{9}, Doornik\cite{10} and Carneiro et al.\cite{11} proved that Google Trends data could be predictive for current influenza-like activity levels by 1-2 weeks earlier before conventional centers for disease control and prevention surveillance systems by comparing GT data and the actual disease numbers and provided different case studies. The search frequency would dramatically increase before and during the disease outbreak. Similarly, Cook et al.\cite{12} chose H1N1 ease cases. The increasing search frequency could be useful in identifying the presence of diseases and the media effect on web users' search behaviors \cite{13}. 

GT data was proven to be effective in terms of modeling other areas such as marketing and information security. Youn et al.\cite{14} used GT data and Autoregressive Integrated Moving Average (ARIMA) models to conduct nowcast for TV market of a few brands and was able reveal the correlation.  Accurate prediction for the near future of the market was obtained. Rech \cite{15} used GT data to analyze the attention that products received and the cause-effect relation among a few factors in software engineering. Kuo et al \cite{16} demonstrated the lifecycles of internet security systems had the same pattern including four stages: zero day, publicity, cooldown and silence with different scales. The author discovered that GT data showed a interesting correlation with the lifecycle and claimed the reason was that when the vulnerability attracted a lot attention, the risk became large and the lifecycle turned to the decay stages. Choi et al \cite{17} was able to conduct time series analysis on GT data to forecast some economic indicators, and showed that some GT data was very well fitted by ARIMA models. In their case, they focused on the intrinsic structure of the data sets without incorporating any explanatory variables or time series. Mondal et al \cite{18} used transfer function noise model to study the effect of monthly rain fall on the Ganges River flow, with both data sets being time series. In our case, we will use an impulse series as the explanatory. 

Ari Seifter\cite{19} show that GT data was high related to the public attention on diseases according to a study on Lyme disease. Grant\cite{20} analyzed the number of articles published and number of early detection of disease in the event month for breast cancer and concluded that the event did promote public attention. The study quantitatively indicated that a successful event actually educated public and encouraged early detection. Here we want to identify the effective ones from a pool of events. In \cite{21}, Ayers et al  studied the Great American Smokeout health awareness event by using a number of data sets such as number of news, tweets, Wiki visits and etc. Their proposed evaluation method for event effectiveness was to first fit counterfactual data by assuming the event had not occurred, then compare them with the actual data. Although their approach was quantitative, they used the percent change where it is unclear detect the threshold of significance.

\newsec{Datasets and Preprocessing}{DP}
\newsub{Datasets}{da}
We focus on monthly health awareness events in the US and select a set of 46 events on disease. Since GT data is based on the search frequency of one or a few words which we call a query, we select a query for each event and present them in Appendix A. In fact, for some events, there were more than one meaningful queries, then we picked the one with highest frequency. 

On Google Trends webpage, users are able to track the search popularity of queries in different languages across regions starting from January 2004. Weekly or monthly GT data may be downloaded as a CSV file depending on the total time range. Since the pure values of queries can be huge numbers, Google rescales them in a range from 0 to 100 with the highest frequency being 100. Four options, including Region, Time, Category and Search Type are needed to specify a search and are selected as United State, 2004-2017, Health, and Web search respectively in this work.

For example, Figure 1 shows the query of Breast Cancer as a time series plot. There was also a graph showing popularity over regions as shown in Figure \ref{stationmapd}. the top three subregions of search popularity were Pennsylvania, Maryland, Alabama.

\begin{figure}[!h]
\begin{center}
\leavevmode
\includegraphics[height=2.8in]{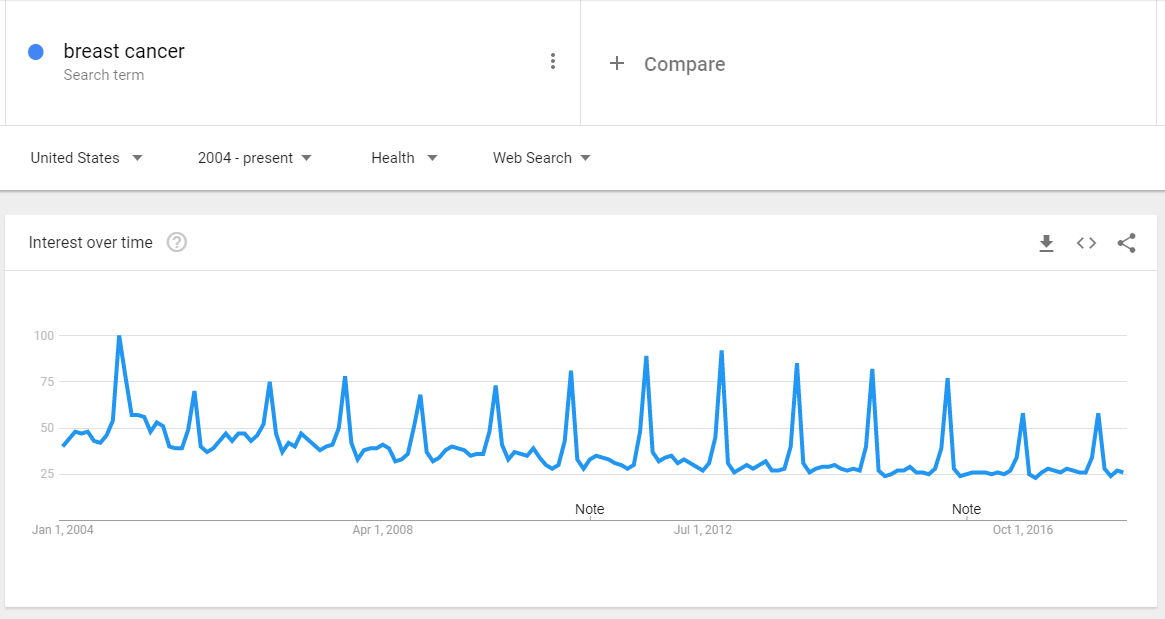}
\end{center}
\vspace{-.1in}
\caption[search engines]{Google Trends Search Plot for the Query of Breast Cancer}
\label{stationmapc}
\end{figure}

\begin{figure}[!h]
\begin{center}
\leavevmode
\includegraphics[height=1.8in]{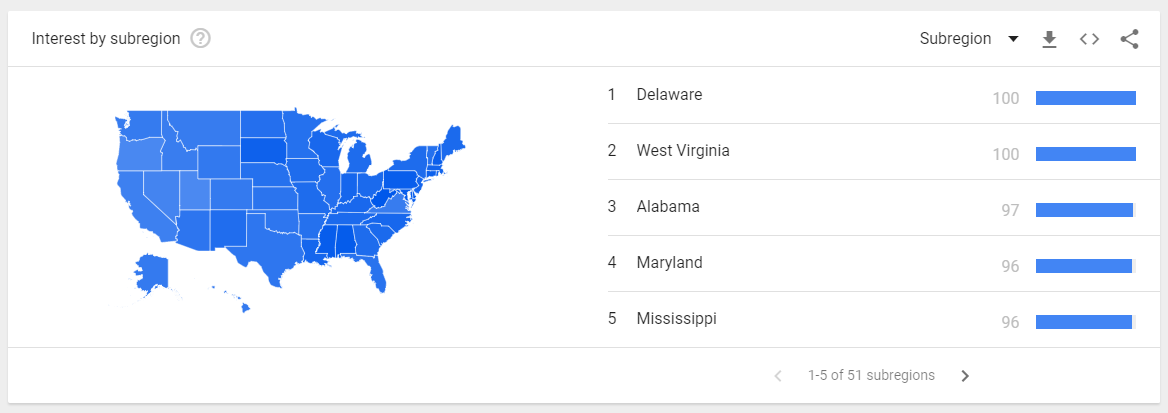}
\end{center}
\vspace{-.1in}
\caption[search engines]{Google Trends Search by Region for the Query of Breast Cancer}
\label{stationmapd}
\end{figure}

\newsub{Data Preprocessing}{dap}
Monthly data from 2004 to 2017 for 46 selected queries are collected. All data points are integers between 0 and 100, with no missing data. We rescale every month to an equal length of 30 days to reduce the variation caused by uneven number of days. In particular, January, March, May, July, August, October, and December data points are multiplied by $\frac{30}{31}$, and February data points are multiplied by $\frac{30}{28}$.

\newsec{Methodology}{me}
We provide three different quantitative methods to evaluate the effectiveness with their thresholds clearly stated. The main method is to use transfer function noise modelling with impulse series as input. Then inferences based on Wilcoxon Rank Sum test and Binomial distribution are used to consolidate results.
\newsub{Transfer Function Noise Model}{xmodel}

The (Seasonal) Autoregressive Integrated Moving Average models (ARIMA or SARIMA) make interpretation and forecast by developing the intrinsic pattern of the single response time series. A general SARIMA $(p,d,q)(P,D,Q)_s$ has the form:

\begin{align*}
(1-\sum_{i=1}^P \phi^{\ast}_i B^{is})(1-\sum_{i=1}^p \phi_iB^i)&(1-B^s)^D(1-B)^dy_t\\
=&(1+\sum_{i=1}^Q \theta^{\ast}_i B^{is})(1+\sum_{i-1}^q\theta_i B^i)\epsilon_t,
\end{align*}
where $B$ is the backshift operator, $By_t=y_{t-1}$, $\epsilon_t$ is a white noise, and $\phi_i$, $\theta_i$, $\phi^{\ast}_i$, and $\theta^{\ast}_i$ are constant coefficients. This model can be expressed by a more compact notation as:

$$\phi(B)y_t=\theta(B)\epsilon_t \iff y_t=\frac{\theta(B)}{\phi(B)}\epsilon_t$$

If there is another series, say $\{x_t\}$ which we call an input series that has a relationship with $\{y_t\}$. The Transfer Function Noise Model is built to describe this situation as 

\begin{equation}
y_t=c+\frac{w(B)B^b}{\delta(B)}x_t+\frac{\theta(B)}{\phi(B)}\epsilon_t
\end{equation}

Intuitively, $\frac{w(B)B^b}{\delta(B)}$ is determined by the structure of input $\{x_t\}$ and measures the effect of $\{x_t\}$ on $\{y_t\}$, and $\frac{\theta(B)}{\phi(B)}\epsilon_t$ measures the intrinsic pattern with $\{y_t\}$ itself.

We construct an impulse time series $\{x_t\}$ with $x_i=0$ if it corresponds a non event month, and $x_i=1$ if it corresponds an event month. We want to analyze the effect of $\{x_t\}$ towards $\{y_t\}$. (3.1) is called the Intervention model, whose operator $\frac{w(B)B^b}{\delta(B)}$ usually has a fairly simple form. We let $\frac{w(B)B^b}{\delta(B)}=w_0$, and we are interested in how much the impulse $\{x_t\}$ contributes to the current response $\{y_t\}$ which results in:

\begin{equation}
y_t=c+w_0x_t+\frac{\theta(B)}{\phi(B)}\epsilon_t
\end{equation}

We first determine whether there is a seasonality in each data set, that is whether an ARIMA model or a SARIMA model should be used and then fit the best ARIMA/SARIMA model.

Secondly, we fit a transfer function noise model. The input series is just impulse function, thus there is no prewhitening step. To determine the orders of $\theta(B)$ and $\phi(B)$ in (3.2), we use two attempts and choose the better one:

The first attempt will be simply to use the same order as the ARIMA/SARIMA. In second attempt, we first replace the event month data with the average of the previous and next month. The idea is that after this replacement, the new data is our best guess for what the data would be if there were no event happening. We use the new data to determine the orders of the ARIMA/SARIMA model and use them in (3.2). The better attempt is chosen as the final transfer function noise model.

We will conclude that the event contributes to the number of search if the transfer function noise model is better fitted than the ARIMA/SARIMA model, and the parameter $w_0$ is significant at $0.05$ level.

\newsub{Wilcoxon Rank Sum Test}{wrst}
The Wilcoxon Rank Sum test was introduced by Frank Wilcoxon in his well-known article \cite{22} to compare the means of two groups. Clifford \cite{23} showed that Wilson test usually holds large power advantages over t test and is asymptotically more efficient than t test.  In our case, the sample sizes are unequal and the sample distributions are unclear, thus we believe the Wilcoxon Rank-Sum is more appropriate than the t-test.


Data points are splitted into two groups as event month and non event month. The question then become that if event-month group has larger values. The null hypothesis is that the two group of observations came from the same population. The Wilcoxon test is based upon ranking data points of the combined sample. Assign numeric ranks to all the observations with 1 being the smallest value. If there is a group that ties, assign the rank equal to its average ranking. The Wilcoxon rank-sum test statistic is the sum of the ranks for observations from one of the samples and therefore are calculated as:
\begin{equation}
U_x=n_xn_y+\frac{n_x(n_x+1)}{2}-u_x
\end{equation}
\begin{equation}
U_y=n_xn_y+\frac{n_y(n_y+1)}{2}-u_y
\end{equation}

where $n_x$ and $n_y$ are the two sample sizes; $u_x$ and $u_y$ are the sums of the ranks in samples $x$ and $y$ respectively. The smaller value between $U_x$  and $U_y$  is the one used to consult significance tables to estimate the p-value.

\newsub{Inference by Binomial Distribution}{iopn}

We used the null hypothesis that the search frequencies were completely random implying that the event did not have effect. Under the null hypothesis, every month has equal probability $1/12$ to be the peak since all selected diseases are not seasonal as an influenza-like illness. Let $k$ be the number of yearly peaks for event-month data in 14 years. Among 14 years, the probability that a certain month appears to be the peak $k$ times is  

$$P(X=k)=\binom{14}{k}(\frac{1}{12})^k(1-\frac{1}{12})^{(14-k)},\ \ \text{where}\ \ X\sim\mathbf{B(14,\frac{1}{12})}$$

In particular, $k=4$ is the largest value making the probability less than 0.05, and $P(X=4)=0.02$. Therefore, that the event month appears to be the peak at least 4 times indicates evidence that the event-month data is significantly different from the other months.

\newsec{Results}{rad}
Health awareness events that show evidence of significance in all three method decribed above will be defined as effective health awareness events. Health awareness events that have insignificant results for all three tests will be defined as ineffective health awareness events. The events with inconsistent results by different methods will be defined as unclear.

Details for two selected events as case study are presented in this chapter. All 46 selected query data have been analyzed and presented in table \ref{46} in Appendix.


\newsub{Case 1: National Breast Cancer Awareness Month}{nbcam}
One out of eight women in the USA are diagnosed with breast cancer \cite{24}, and breast cancer is the top cause of cancer death for women 40 to 50 years of age \cite{25} and the second leading cause of cancer death for women in the USA \cite{26}. The National Breast Cancer Awareness Event is dedicated to drawing public attention on prevention and early detection, supporting the patients and fundraising for scientific research.



\begin{figure}[!h]
\begin{center}
\leavevmode
\includegraphics[height=3.5in]{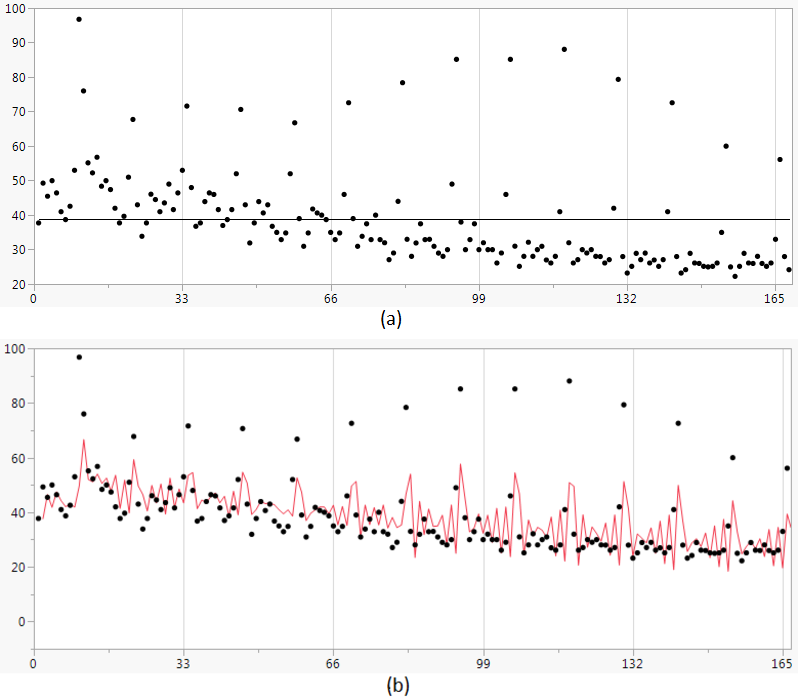}
\end{center}
\vspace{-.3in}
\caption[search engines]{Breast Cancer: (a) shows a Time Series Plot; (b) shows the fitted ARIMA line.}
\label{stationmap10}
\end{figure}

The time series plot as shown in Figure \ref{stationmap10} presented a slightly declining trend, with peaks at the event months, October. Three different tests including periodogram, auto-correlation function, and linear model comparison are conducted to check the seasonality. For breast cancer data, two of the three tests indicated that there is no seasonality, therefore we choose ARIMA model instead of SARIMA and obtain the best ARIMA model and transfer function model as described in section 4.1.

The results are shown in table \ref{tablecase1}. We see that the Adjust $R^2$ is about 0.41 for the ARIMA model and is about 0.58 for the transfer function noise model, and the p-value for $\{x_t\}$ parameter ``eventmonth" is $<0.0001$. Therefore we conclude that the event has a significant effect on the number of search for breast cancer.

\begin{table}[]
\centering

\begin{tabular}{|l|l|l|l|}
\hline
       & Orders  & Adjusted R square & p value of event coefficient \\ \hline
ARIMA  & (2,1,3) & 0.408             & NA                           \\ \hline
ARIMAX & (2,0,3) & 0.583             & \textless0.001               \\ \hline
\end{tabular}
\caption{Results for ARIMA and Transfer Function Model(ARIMAX)}
\label{tablecase1}
\end{table}



Next, to conduct the Wilcoxon rank sum test, we split the data into event month subset and non event month subset. A  p-value $0.0000 < 0.05$ indicate that we shall reject the null hypothesis that two groups of observations come from the same population. Further we notice that the mean of the event months is greater than non event months, thus during event months the search frequencies are higher than the rest of the year.

\begin{figure}[!h]
\begin{center}
\leavevmode
\includegraphics[height=2.6in]{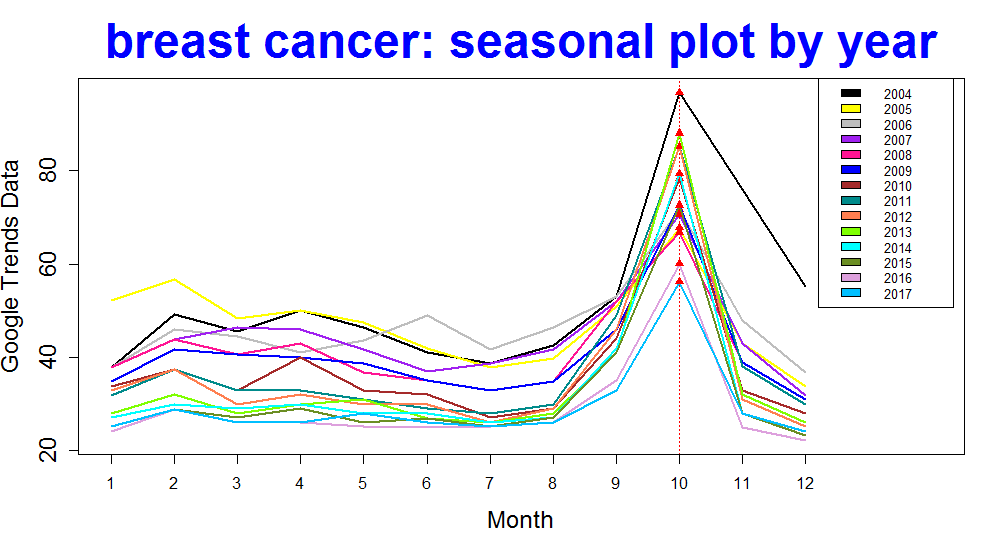}
\end{center}
\vspace{-.4in}
\caption[search engines]{Breast Cancer: All 14 Peaks Fall in October.}
\label{stationmap13}
\end{figure}

For the Binomial approach, among 14 years of Google Trends data of the query breast cancer, we have found that all 14 yearly peaks happen in October(see Color Figure \ref{stationmap13}) which is greater than the threshold, 4. There is evidence to prove that event-month frequencies are greater than the other months'. 

In sum, all our results consistently indicate that the National Breast Cancer Awareness event is effective in increasing search frequency of breast cancer in October.

\newsub{Case 2: American Stroke Awareness Month}{asam}
Strokes are one of the leading causes of death and serious long-term disability in the USA \cite{27}. More than 795,000 Americans have a stroke every year and about 130,000 people have been killed by a stroke in the USA each year \cite{28}. To get insight into public awareness for American Stroke Awareness Month, Google Trends data of query stroke has been obtained.






\begin{figure}[!h]
\begin{center}
\leavevmode
\includegraphics[height=3.5in]{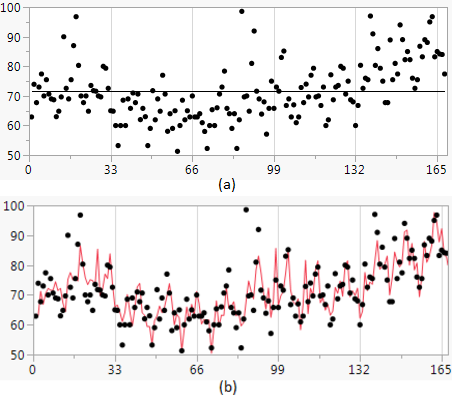}
\end{center}
\vspace{-.3in}
\caption[search engines]{Stroke: (a) shows a Time Series Plot; (b) shows the fitted SARIMA line.}
\label{stationmap30}
\end{figure}

The time series plot as shown in Figure \ref{stationmap30} (a) presents a slight decline trend before the year 2011 and an uptrend after the year 2011. We use a R function which uses three different tests including peridogram, auto-correlation function, and linear model comparison to check the seasonality. For stroke data, all three tests indicate that there is seasonality, meaning SARIMA model should be used. The outputs for SARIMA model and transfer function noise model are presented in Figure \ref{tablecase3}. We see that the Adjust $R^2$ is about 0.62 for the transfer function noise model which is no better than the one for SARIMA model which is about 0.68, and the p-value for $\{x_t\}$ parameter ``eventmonth" is about $0.235>0.05$. Therefore we do not have evidence to conclude that the event has a significant effect on the number of search for Stroke.

\begin{table}[]
\centering
\begin{tabular}{|l|l|l|l|}
\hline
       & Orders         & Adjusted R square & p value of event coefficient \\ \hline
SARIMA & (4,1,2)(2,0,0) & 0.677             & NA                           \\ \hline
ARIMAX & (4,1,2)(2,0,0) & 0.620             & 0.2354                       \\ \hline
\end{tabular}
\caption{Results for ARIMA and Transfer Function Model(ARIMAX)}
\label{tablecase3}
\end{table}



According to the one-side Wilcoxon Rank-Sum test statistics, we have p-value$=0.2918>0.05$, which means we have no compelling evidence that there is higher search frequency for the query “strokes” in the event month of May.

\begin{figure}[!h]
\begin{center}
\leavevmode
\includegraphics[height=2.6in]{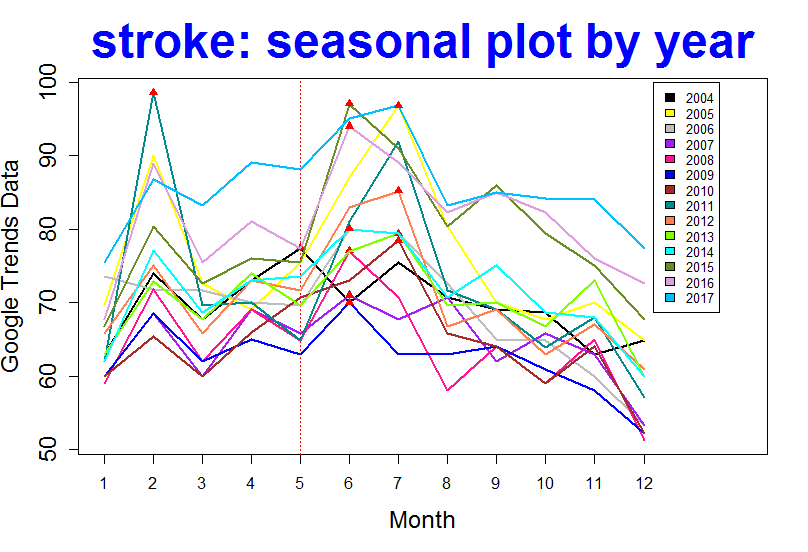}
\end{center}
\vspace{-.4in}
\caption[search engines]{Stroke: one peak falls in May.}
\label{stationmap33}
\end{figure}

From the years 2004 to 2017, we have only one peak in May (See color Figure \ref{stationmap33}) which is less than the threshold of four peaks. In sum, all our results consistently indicate that the there is no evidence that the Stroke Awareness event is effective in increasing search frequency of stroke in May.



Ten events are concluded to be effective in raising public search frequency about related diseases: Alcohol Awareness, Autism, Breast Cancer, Colon Cancer, Dental Health, Heart Disease, Immunization, National Nutrition, Ovarian Cancer, and Sids. Eight events are unclear according to inconsistent results and the others are ineffective.

\newsec{Conclusion and Discussion}{concl}
According to the analysis of all 46 data sets, we have found that 10 health awareness events are effective health awareness events by showing strong evidence of significant seasonal patterns with peaks matching the event month, 28 events are defined as ineffective health awareness events and the rest are defined as unclear health awareness events. Although lack of attention is definitely bad, overheating events may result in possessing too much public resources and weakening the severity of other health topics.





People may suspect that the effective events should have higher frequencies than others, or the opposite. In fact, we checked the relative frequencies across effective, unclear, and ineffective events, and found that there is no relationship. There are effective events with high search frequencies and low search frequencies, and vice versa. 

Another interesting thing to notice is that Diabetes was classified as unclear, which was somehow counterintuitive. We compared all eight unclear events and found out that the frequency for Diabetes is absolutely the largest, while all other 7 events are relatively closed to each other but away from Diabetes.  We suspected that the Diabetes is so influential that a considerable attention was paid on it during many months over a year which made the event month insignificant. Therefore, a possible future study is to think about if some of the unclear and inffective events are similar to the case of Diabetes. We may also consider the prevalence and severity of these disease, since obviously it is not practical to make all disease as well-known as heart disease or breast cancer.


Classification within this study will be beneficial for the public health management and health awareness for public welfare. The Department of Public Health and public interest groups need to optimally rearrange resources allocation between effective health awareness events and ineffective health awareness events to improve the awareness of ineffective health awareness events topics, especially. Corporate partners would take the opportunity to promote related products or services to effective health awareness events, such as pink-ribbon brooch, exclusive pink-ribbon products, and the clinic needs to be prepared for increased demands of health screening appointments.
 

\bibliographystyle{abbrv}
\bibliography{simple}

\newpage
\appendix
\section*{Appendices}
\renewcommand{\thesubsection}{\Alph{subsection}}
\subsection{National Health Awareness Events with corresponding Selected Queries}
\begin{table}[!h]
\begin{tabular}{|l|l|}
\hline
Health Awareness Event/Month                 & Query               \\ \hline
January                                      &                     \\ \hline
National Birth Defects Prevention Month      & Birth Defects       \\ \hline
Cervical Health Awareness Month              & Cervical            \\ \hline
National Glaucoma Awareness Month            & Glaucoma            \\ \hline
Thyroid Awareness Month                      & Thyroid             \\ \hline
                                             &                     \\ \hline
February                                     &                     \\ \hline
American Heart Month                         & Heart Disease       \\ \hline
National Children's Dental Health Month      & Dental Health       \\ \hline
                                             &                     \\ \hline
March                                        &                     \\ \hline
National Colorectal Cancer Awareness Month   & Colon Cancer        \\ \hline
National Endometriosis Awareness Month       & Endometriosis       \\ \hline
National Nutrition Month                     & National Nutrition  \\ \hline
Multiple Sclerosis Education Month           & Sclerosis           \\ \hline
                                             &                     \\ \hline
April                                        &                     \\ \hline
Alcohol Awareness Month                      & Alcohol Awareness   \\ \hline
National Autism Awareness Month              & Autism              \\ \hline
Irritable Bowel Syndrome Month               & Ibs                 \\ \hline
                                             &                     \\ \hline
May                                          &                     \\ \hline
American Stroke Awareness Month              & Stroke              \\ \hline
Arthritis Awareness Month                    & Arthritis           \\ \hline
National Asthma and Allergy Awareness Month  & Asthma Allergy      \\ \hline
National Celiac Disease Awareness Month      & Celiac              \\ \hline
Hepatitis Awareness Month                    & Hepatitis           \\ \hline
National High Blood Pressure Education Month & High Blood Pressure \\ \hline
Lupus Awareness Month                        & Lupus               \\ \hline
Mental Health Month                          & Mental Health       \\ \hline
National Osteoporosis Awareness Month        & Osteoporosis        \\ \hline
Skin Cancer Detection and Prevention Month   & Skin Cancer         \\ \hline
\end{tabular}
\end{table}

\newpage
\begin{table}[!h]
\begin{tabular}{ll}
\hline
\multicolumn{1}{|l|}{Health Awareness Event/Month}                          & \multicolumn{1}{l|}{Query}                 \\ \hline
\multicolumn{1}{|l|}{June}                                                  & \multicolumn{1}{l|}{}                      \\ \hline
\multicolumn{1}{|l|}{National Aphasia Awareness Month}                      & \multicolumn{1}{l|}{aphasia}               \\ \hline
\multicolumn{1}{|l|}{Scoliosis Awareness Month}                             & \multicolumn{1}{l|}{scoliosis}             \\ \hline
\multicolumn{1}{|l|}{}                                                      & \multicolumn{1}{l|}{}                      \\ \hline
\multicolumn{1}{|l|}{July}                                                  & \multicolumn{1}{l|}{}                      \\ \hline
\multicolumn{1}{|l|}{Eye Injury Prevention Month}                           & \multicolumn{1}{l|}{eye injury}            \\ \hline
\multicolumn{1}{|l|}{}                                                      & \multicolumn{1}{l|}{}                      \\ \hline
\multicolumn{1}{|l|}{August}                                                & \multicolumn{1}{l|}{}                      \\ \hline
\multicolumn{1}{|l|}{Amblyopia Awareness Month}                             & \multicolumn{1}{l|}{amblyopia}             \\ \hline
\multicolumn{1}{|l|}{National Immunization Awareness Month}                 & \multicolumn{1}{l|}{immunization}          \\ \hline
\multicolumn{1}{|l|}{Psoriasis Awareness Month}                             & \multicolumn{1}{l|}{psoriasis}             \\ \hline
\multicolumn{1}{|l|}{}                                                      & \multicolumn{1}{l|}{}                      \\ \hline
\multicolumn{1}{|l|}{September}                                             & \multicolumn{1}{l|}{}                      \\ \hline
\multicolumn{1}{|l|}{National Alcohol and Drug Addition Recovery Month}     & \multicolumn{1}{l|}{alcohol drug addition} \\ \hline
\multicolumn{1}{|l|}{National Cholesterol Education Month}                  & \multicolumn{1}{l|}{cholesterol}           \\ \hline
\multicolumn{1}{|l|}{Lcukemia and Lymphomn Awareness Month}                 & \multicolumn{1}{l|}{Lcukemia}              \\ \hline
\multicolumn{1}{|l|}{National Menopause Awareness Month}                    & \multicolumn{1}{l|}{menopause}             \\ \hline
\multicolumn{1}{|l|}{Ovarian Cancer Awareness Month}                        & \multicolumn{1}{l|}{ovarian cancer}        \\ \hline
\multicolumn{1}{|l|}{Prostate Awareness Month}                              & \multicolumn{1}{l|}{prostate}              \\ \hline
\multicolumn{1}{|l|}{}                                                      & \multicolumn{1}{l|}{}                      \\ \hline
\multicolumn{1}{|l|}{October}                                               & \multicolumn{1}{l|}{}                      \\ \hline
\multicolumn{1}{|l|}{National Breast Cancer Awareness Month}                & \multicolumn{1}{l|}{breast cancer}         \\ \hline
\multicolumn{1}{|l|}{National Dental Hygiene Month}                         & \multicolumn{1}{l|}{dental hygiene}        \\ \hline
\multicolumn{1}{|l|}{National Depression and Mental Health Screening Month} & \multicolumn{1}{l|}{depression}            \\ \hline
\multicolumn{1}{|l|}{National Down Syndrome Awareness Month}                & \multicolumn{1}{l|}{down syndrome}         \\ \hline
\multicolumn{1}{|l|}{SIDS Awareness Month}                                  & \multicolumn{1}{l|}{Sids}                  \\ \hline
\multicolumn{1}{|l|}{Spina Bifida Awareness Month}                          & \multicolumn{1}{l|}{spina bifida}          \\ \hline
\multicolumn{1}{|l|}{}                                                      & \multicolumn{1}{l|}{}                      \\ \hline
\multicolumn{1}{|l|}{November}                                              & \multicolumn{1}{l|}{}                      \\ \hline
\multicolumn{1}{|l|}{National Alzheimer's Disease Awareness Month}          & \multicolumn{1}{l|}{alzheimer}             \\ \hline
\multicolumn{1}{|l|}{American Diabetes Month}                               & \multicolumn{1}{l|}{diabetes}              \\ \hline
\multicolumn{1}{|l|}{National Epilepsy Awareness Month}                     & \multicolumn{1}{l|}{epilepsy}              \\ \hline
\multicolumn{1}{|l|}{Lung Cancer Awareness Month}                     & \multicolumn{1}{l|}{ lung cancer}              \\ \hline
\multicolumn{1}{|l|}{Pancreatic Cancer Awareness Month}                     & \multicolumn{1}{l|}{pancreatic cancer }              \\ \hline                  
\end{tabular}
\end{table}

\newpage



\subsection{The results of three methods for all 46 query data}

\begin{table}[!h]
\footnotesize{
\centering
\begin{tabular}{llllll}
Event                                                            & \begin{tabular}[c]{@{}l@{}}Wilcox \\ Sum Test\\ p-value\end{tabular} & \begin{tabular}[c]{@{}l@{}}Peaks at \\ Event \\ Months\end{tabular} & \begin{tabular}[c]{@{}l@{}}Transfer Function\\ Noise Model \\ Fits Better\end{tabular} & \begin{tabular}[c]{@{}l@{}}Input Series \\ Coefficient\\ p value\end{tabular} & Conclusion  \\
                                                                 &                                                                      &                                                                     &                                                                                        &                                                                               &             \\
Alcohol Awareness                                                & 0.0013*                                                              & 6                                                                   & Yes                                                                                    & 0*                                                                            & Effective   \\
Autism                                                           & 0*                                                                   & 12                                                                  & Yes                                                                                    & 0*                                                                            & Effective   \\
Breast Cancer                                                    & 0*                                                                   & 14                                                                  & Yes                                                                                    & 0*                                                                            & Effective   \\
Coloncancer                                                      & 0.0008*                                                              & 7                                                                   & Yes                                                                                    & 0.0129*                                                                       & Effective   \\
Dental Health                                                    & 0*                                                                   & 14                                                                  & Yes                                                                                    & 0*                                                                            & Effective   \\
Heart Disease                                                    & 0*                                                                   & 14                                                                  & Yes                                                                                    & 0.0016*                                                                       & Effective   \\
Immunization                                                     & 0*                                                                   & 14                                                                  & Yes                                                                                    & 0.0009*                                                                       & Effective   \\
National Nutrition                                               & 0*                                                                   & 5                                                                   & Yes                                                                                    & 0.0054*                                                                       & Effective   \\
Ovarian Cancer                                                   & 0.0007*                                                              & 7                                                                   & Yes                                                                                    & 0*                                                                            & Effective   \\
Sids                                                             & 0.0008*                                                              & 4                                                                   & Yes                                                                                    & 0*                                                                            & Effective   \\
Asthma Allergy                                                   & 0.0183*                                                              & 3                                                                   & Yes                                                                                    & 0.0636                                                                        & Unclear     \\
Diabetes                                                         & 0.0297*                                                              & 1                                                                   & No                                                                                     & 0.0813                                                                        & Unclear     \\
Endometriosis                                                    & 0.1314                                                               & 4                                                                   & No                                                                                     & 0.7099                                                                        & Unclear     \\
Epilepsy                                                         & 0.0159*                                                              & 0                                                                   & No                                                                                     & 0.2426                                                                        & Unclear     \\
Lung Cancer                                                      & 0.0341*                                                              & 1                                                                   & No                                                                                     & 0.1929                                                                        & Unclear     \\
Lupus                                                            & 0.0192*                                                              & 4                                                                   & Yes                                                                                    & 0.7506                                                                        & Unclear     \\
Menopause                                                        & 0.0177*                                                              & 2                                                                   & No                                                                                     & 0.5078                                                                        & Unclear     \\
Skin Cancer                                                      & 0                                                                    & 5                                                                   & No                                                                                     & 0.0504                                                                        & Unclear     \\
\begin{tabular}[c]{@{}l@{}}Alcohol Drug\\ Addiction\end{tabular} & 0.3959                                                               & 0                                                                   & Yes                                                                                    & 0.0718                                                                        & Ineffective \\
Alzheimer                                                        & 0.177                                                                & 1                                                                   & No                                                                                     & 0.2090                                                                        & Ineffective \\
Amblyopia                                                        & 0.8139                                                               & 1                                                                   & No                                                                                     & 0.9164                                                                        & Ineffective \\
Aphasia                                                          & 0.9809                                                               & 0                                                                   & No                                                                                     & 0.0009*                                                                       & Ineffective \\
Arthritis                                                        & 0.1718                                                               & 1                                                                   & No                                                                                     & 0.6986                                                                        & Ineffective \\
Birth Defect                                                     & 0.1899                                                               & 0                                                                   & No                                                                                     & 0.5783                                                                        & Ineffective \\
Celiac                                                           & 0.22                                                                 & 1                                                                   & No                                                                                     & 0.7075                                                                        & Ineffective \\
Cervical                                                         & 0.8439                                                               & 0                                                                   & No                                                                                     & 0.0012*                                                                       & Ineffective \\
Cholesterol                                                      & 0.2667                                                               & 1                                                                   & No                                                                                     & 0.0124*                                                                       & Ineffective \\
Dental Hygiene                                                   & 0.0724                                                               & 1                                                                   & No                                                                                     & 0.5741                                                                        & Ineffective \\
Depression                                                       & 0.1168                                                               & 1                                                                   & No                                                                                     & 0*                                                                            & Ineffective \\
Down Syndrome                                                    & 0.2446                                                               & 1                                                                   & No                                                                                     & 0.0484*                                                                       & Ineffective \\
Eye Injury                                                       & 0.4793                                                               & 0                                                                   & Yes                                                                                    & 0.2093                                                                        & Ineffective \\
Glaucoma                                                         & 0.6872                                                               & 0                                                                   & Yes                                                                                    & 0.0274*                                                                       & Ineffective \\
Hepatitis                                                        & 0.3914                                                               & 0                                                                   & Yes                                                                                    & 0.0300*                                                                       & Ineffective \\
High Blood Pressure                                              & 0.8289                                                               & 0                                                                   & No                                                                                     & 0.0038*                                                                       & Ineffective \\
Ibs                                                              & 0.1389                                                               & 1                                                                   & No                                                                                     & 0.0033*                                                                       & Ineffective \\
Leukemia                                                         & 0.249                                                                & 0                                                                   & No                                                                                     & 0.0024*                                                                       & Ineffective \\
Mental Health                                                    & 0.5126                                                               & 0                                                                   & No                                                                                     & 0*                                                                            & Ineffective \\
Osteoporosis                                                     & 0.6779                                                               & 0                                                                   & No                                                                                     & 0.0429*                                                                       & Ineffective \\
Pancreatic Cancer                                                & 0.2508                                                               & 0                                                                   & Yes                                                                                    & 0.6771                                                                        & Ineffective \\
Prostate                                                         & 0.7092                                                               & 0                                                                   & No                                                                                     & 0.6659                                                                        & Ineffective \\
Psoriasis                                                        & 0.8311                                                               & 0                                                                   & No                                                                                     & 0.3862                                                                        & Ineffective \\
Sclerosis                                                        & 0.1822                                                               & 0                                                                   & No                                                                                     & 0.0258*                                                                       & Ineffective \\
\end{tabular}}
\label{46}
\end{table}
\newpage

\begin{table}[!h]
\footnotesize{
\centering
\begin{tabular}{llllll}
Event                                                            & \begin{tabular}[c]{@{}l@{}}Wilcox \\ Sum Test\\ p-value\end{tabular} & \begin{tabular}[c]{@{}l@{}}Peaks at \\ Event \\ Months\end{tabular} & \begin{tabular}[c]{@{}l@{}}Transfer Function\\ Noise Model \\ Fits Better\end{tabular} & \begin{tabular}[c]{@{}l@{}}Input Series \\ Coefficient\\ p value\end{tabular} & Conclusion  \\
                                                                 &                                                                      &                                                                     &                                                                                        &                                                                               &             \\
Scoliosis                                                        & 0.3892                                                               & 1                                                                   & Yes                                                                                    & 0.4533                                                                        & Ineffective \\
Spina Bifida                                                     & 0.0036*                                                              & 2                                                                   & No                                                                                     & 0.0047*                                                                       & Ineffective \\
Stroke                                                           & 0.2918                                                               & 1                                                                   & No                                                                                     & 0.2082                                                                        & Ineffective \\
Thyroid                                                          & 0.9551                                                               & 0                                                                   & No                                                                                     & 0.5111                                                                        & Ineffective
\end{tabular}}
\label{46}
\end{table}

\newpage
\subsection{ARIMA and Transfer model comparison in JMP for all 46 events}
Each pair of pictures shows one event, with the left being ARIMA/SARIMA and right one being Transfer Function model.
\begin{figure}[!h]
\begin{center}
\leavevmode
\includegraphics[height=1in]{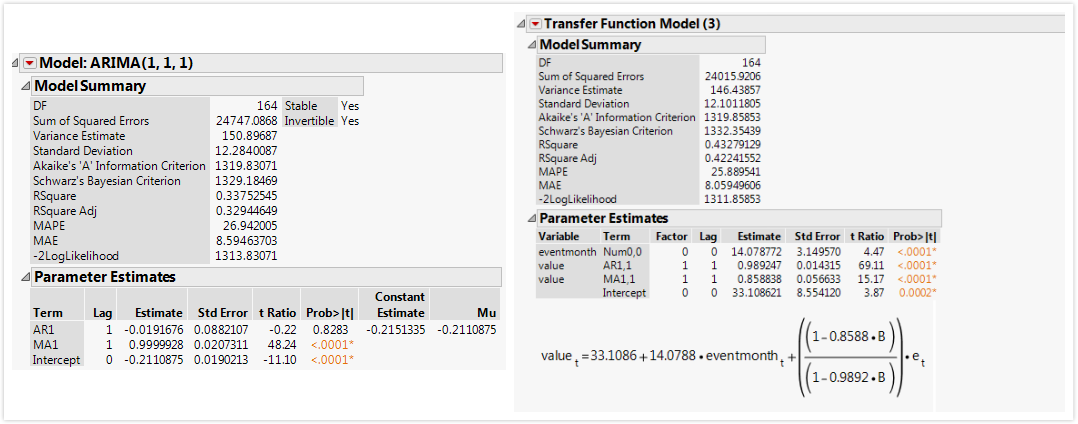}
\end{center}
\vspace{-.1in}
\caption[search engines]{Alcohol Awareness}
\label{stationmapb}
\end{figure}

\begin{figure}[!h]
\begin{center}
\leavevmode
\includegraphics[height=1in]{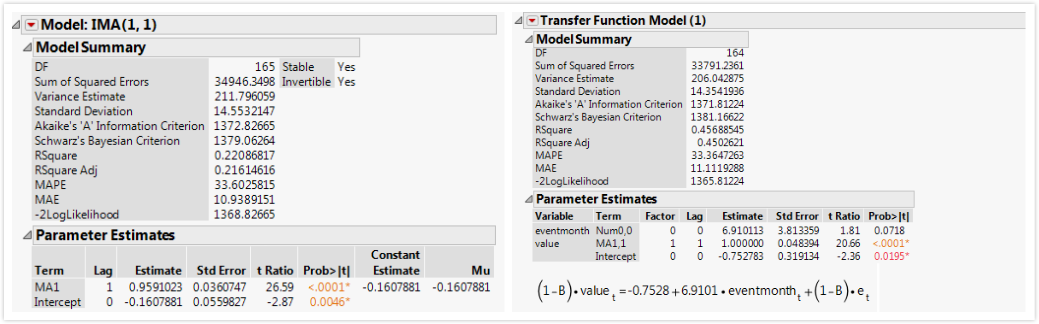}
\end{center}
\vspace{-.1in}
\caption[search engines]{Alcohol Drug Addiction}
\label{stationmapb}
\end{figure}

\begin{figure}[!h]
\begin{center}
\leavevmode
\includegraphics[height=1in]{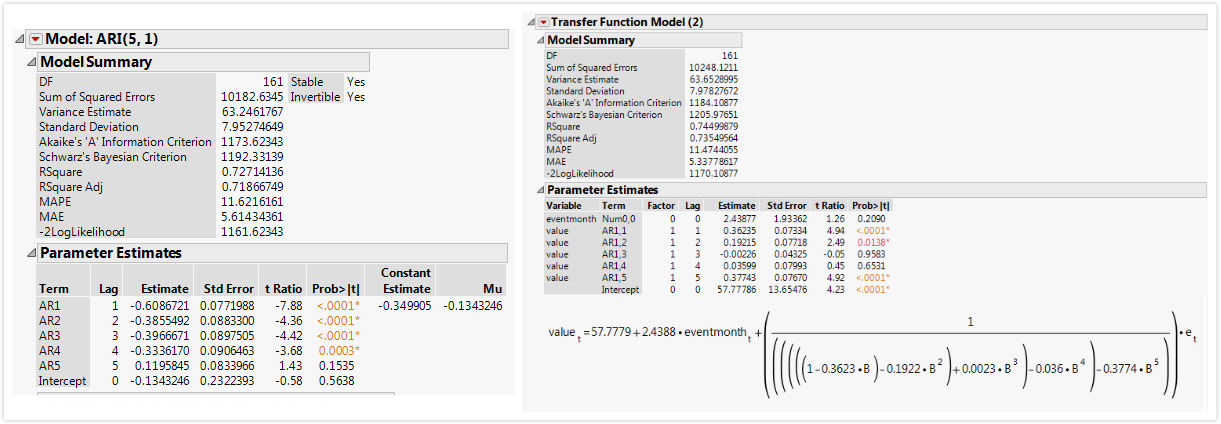}
\end{center}
\vspace{-.1in}
\caption[search engines]{Alzheimer}
\label{stationmapb}
\end{figure}

\begin{figure}[!h]
\begin{center}
\leavevmode
\includegraphics[height=1in]{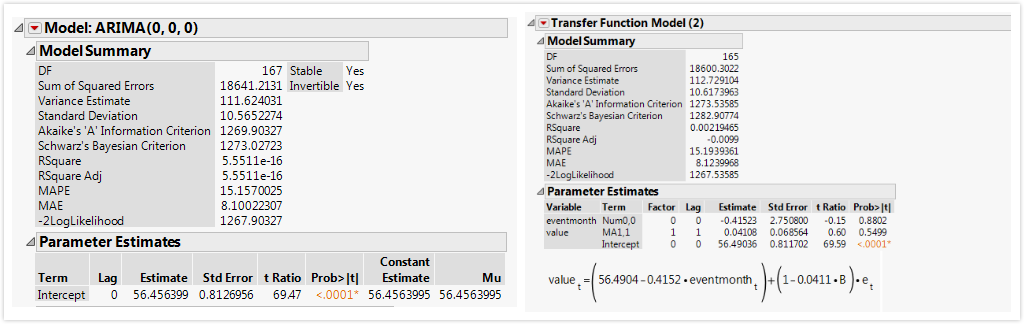}
\end{center}
\vspace{-.1in}
\caption[search engines]{Amblyopia}
\label{stationmapb}
\end{figure}

\begin{figure}[!h]
\begin{center}
\leavevmode
\includegraphics[height=1in]{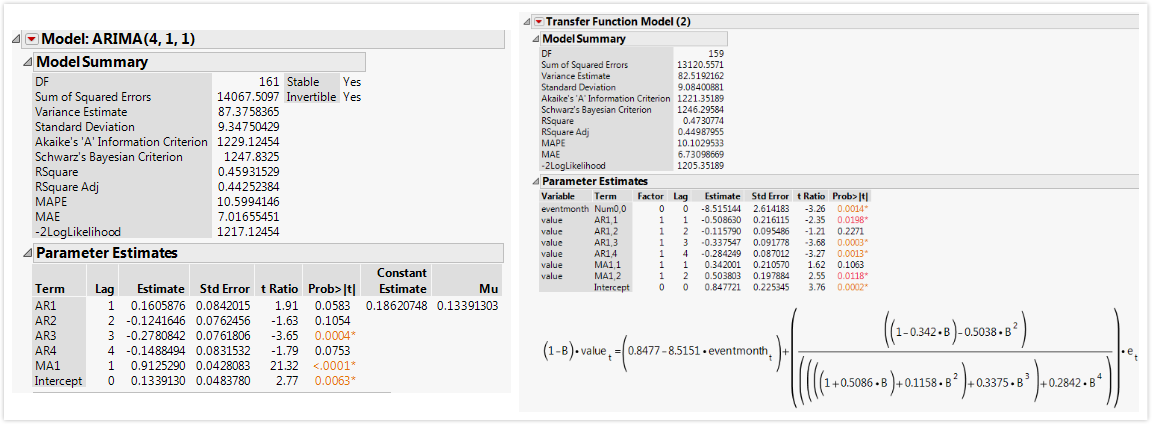}
\end{center}
\vspace{-.1in}
\caption[search engines]{Aphasia}
\label{stationmapb}
\end{figure}

\begin{figure}[!h]
\begin{center}
\leavevmode
\includegraphics[height=1in]{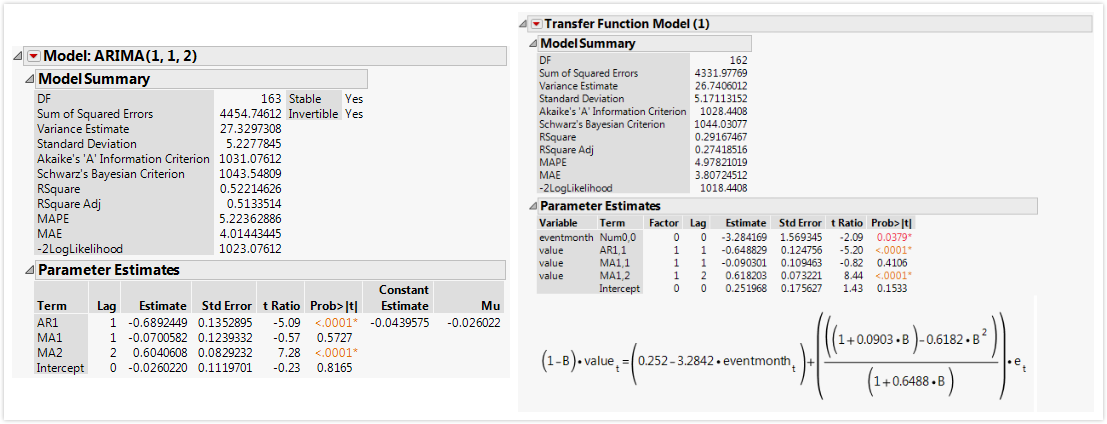}
\end{center}
\vspace{-.1in}
\caption[search engines]{Arthritis}
\label{stationmapb}
\end{figure}

\begin{figure}[!h]
\begin{center}
\leavevmode
\includegraphics[height=1in]{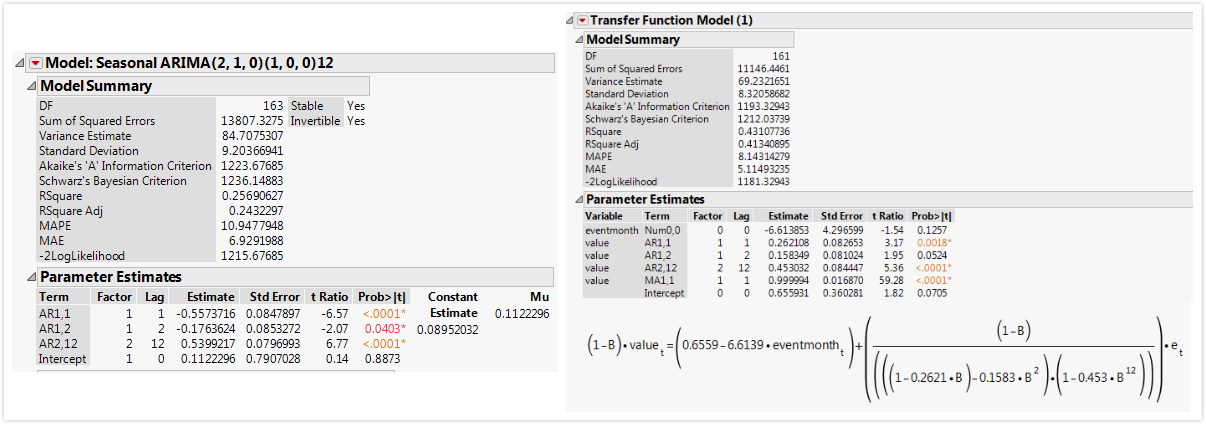}
\end{center}
\vspace{-.1in}
\caption[search engines]{Asthma Allergy}
\label{stationmapb}
\end{figure}

\begin{figure}[!h]
\begin{center}
\leavevmode
\includegraphics[height=1in]{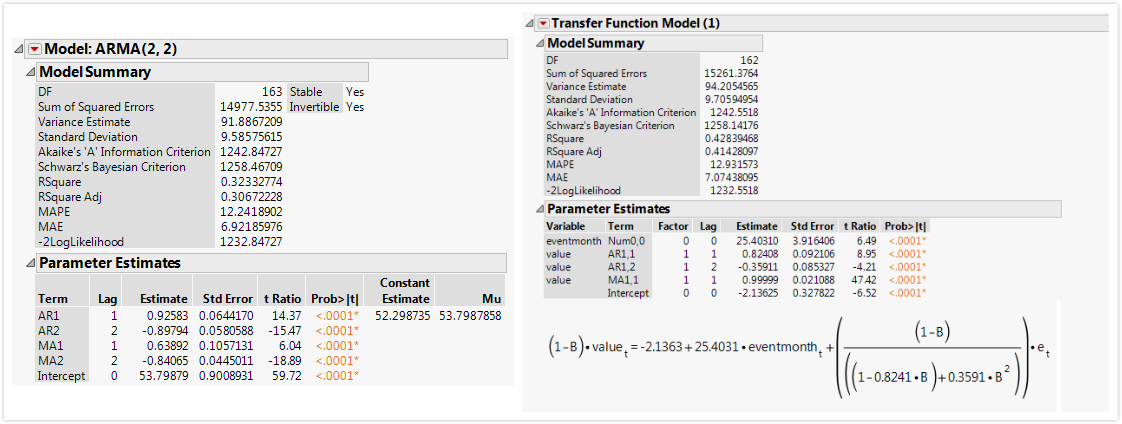}
\end{center}
\vspace{-.1in}
\caption[search engines]{Autism}
\label{stationmapb}
\end{figure}

\begin{figure}[!h]
\begin{center}
\leavevmode
\includegraphics[height=1in]{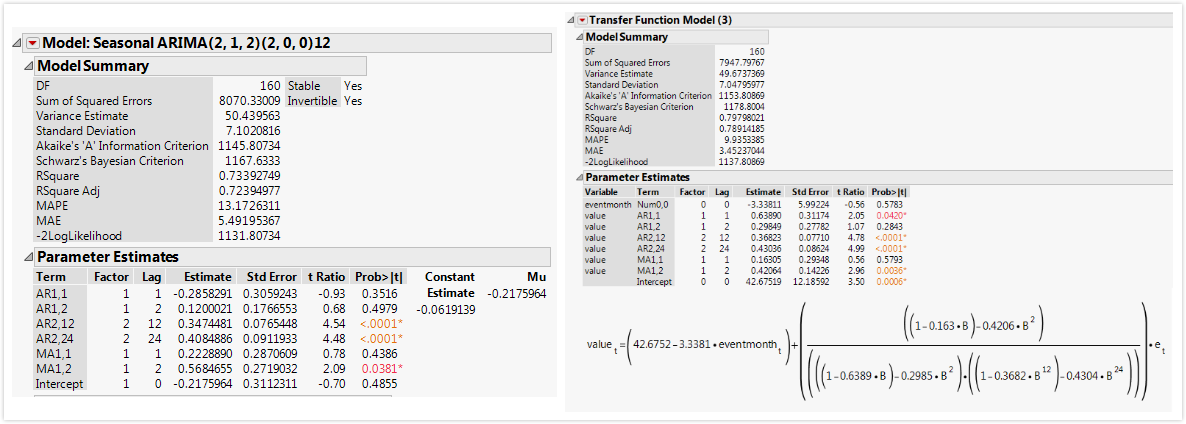}
\end{center}
\vspace{-.1in}
\caption[search engines]{Birth Defect}
\label{stationmapb}
\end{figure}

\begin{figure}[!h]
\begin{center}
\leavevmode
\includegraphics[height=1in]{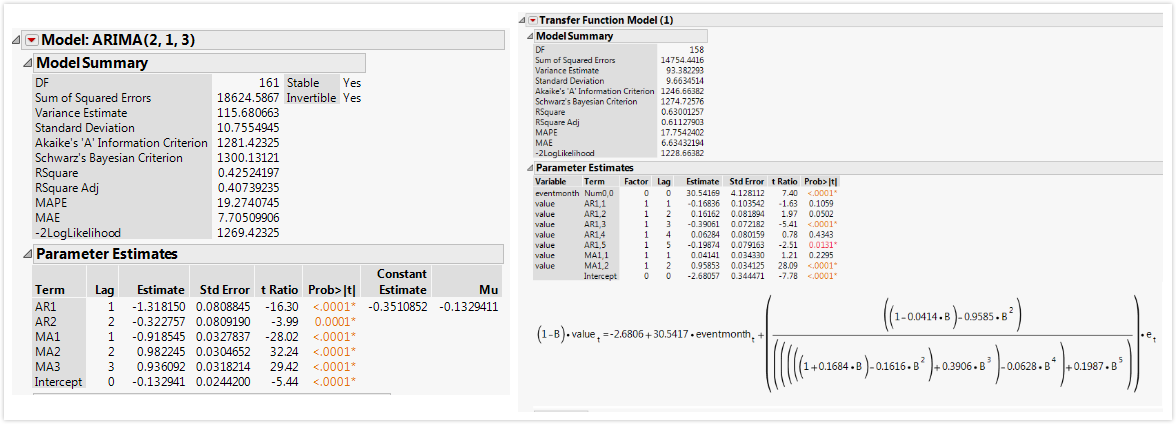}
\end{center}
\vspace{-.1in}
\caption[search engines]{Breast Cancer}
\label{stationmapb}
\end{figure}

\begin{figure}[!h]
\begin{center}
\leavevmode
\includegraphics[height=1in]{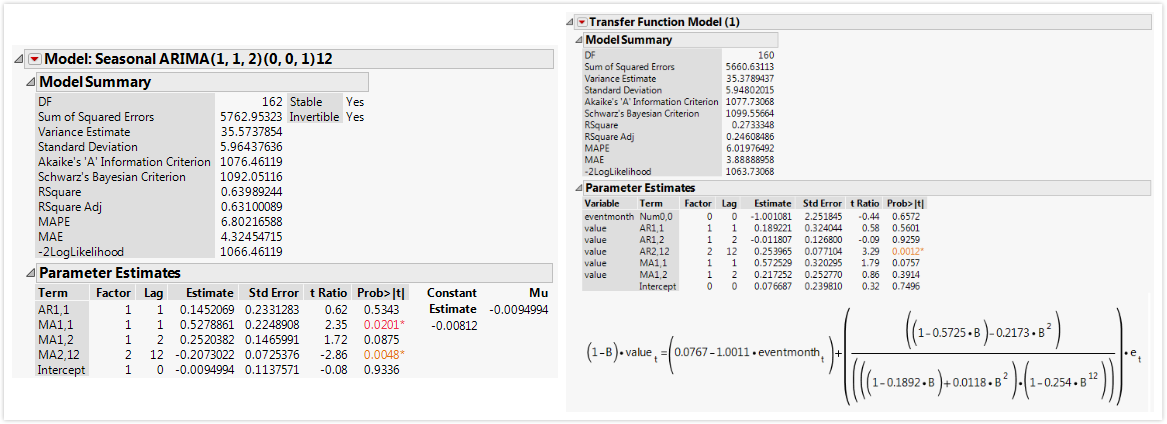}
\end{center}
\vspace{-.1in}
\caption[search engines]{Celiac}
\label{stationmapb}
\end{figure}

\begin{figure}[!h]
\begin{center}
\leavevmode
\includegraphics[height=1in]{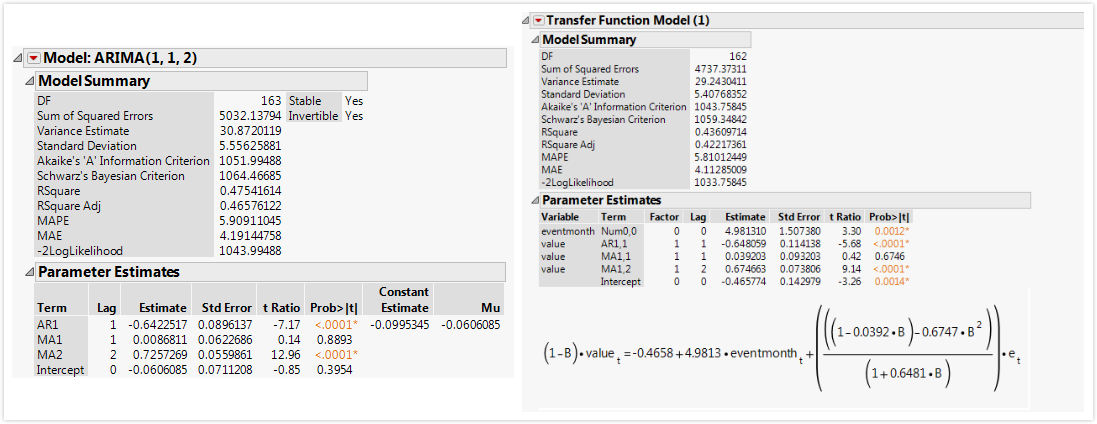}
\end{center}
\vspace{-.1in}
\caption[search engines]{Cervical}
\label{stationmapb}
\end{figure}

\begin{figure}[!h]
\begin{center}
\leavevmode
\includegraphics[height=1in]{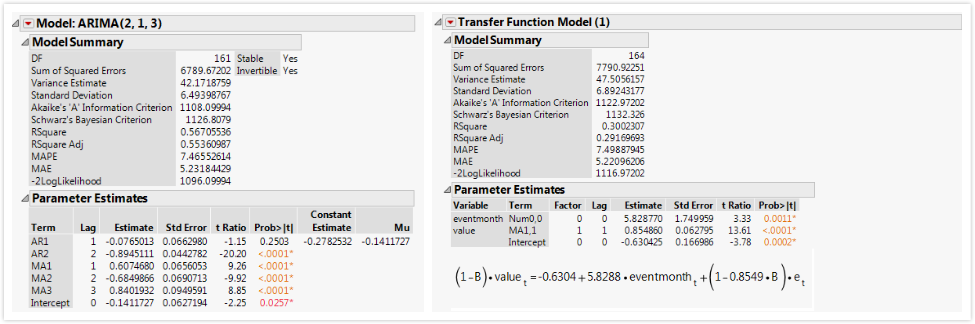}
\end{center}
\vspace{-.1in}
\caption[search engines]{Cholesterol}
\label{stationmapb}
\end{figure}

\begin{figure}[!h]
\begin{center}
\leavevmode
\includegraphics[height=1in]{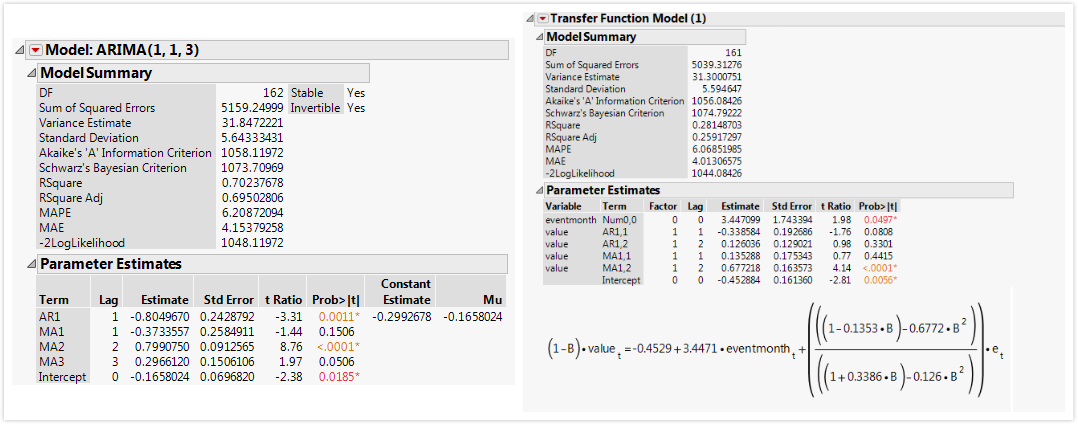}
\end{center}
\vspace{-.1in}
\caption[search engines]{Coloncancer}
\label{stationmapb}
\end{figure}

\begin{figure}[!h]
\begin{center}
\leavevmode
\includegraphics[height=1in]{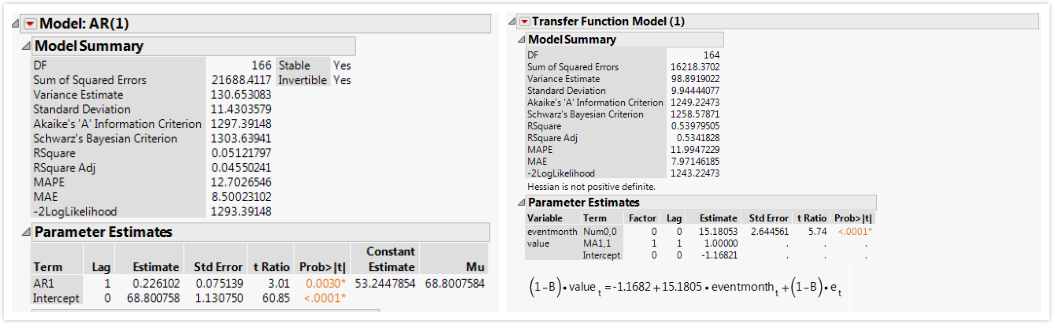}
\end{center}
\vspace{-.1in}
\caption[search engines]{Dental Health}
\label{stationmapb}
\end{figure}

\begin{figure}[!h]
\begin{center}
\leavevmode
\includegraphics[height=1in]{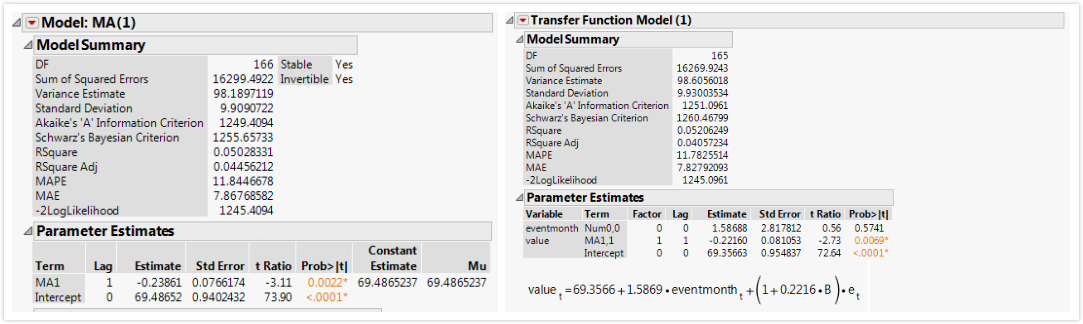}
\end{center}
\vspace{-.1in}
\caption[search engines]{Dental Hygiene}
\label{stationmapb}
\end{figure}

\begin{figure}[!h]
\begin{center}
\leavevmode
\includegraphics[height=1in]{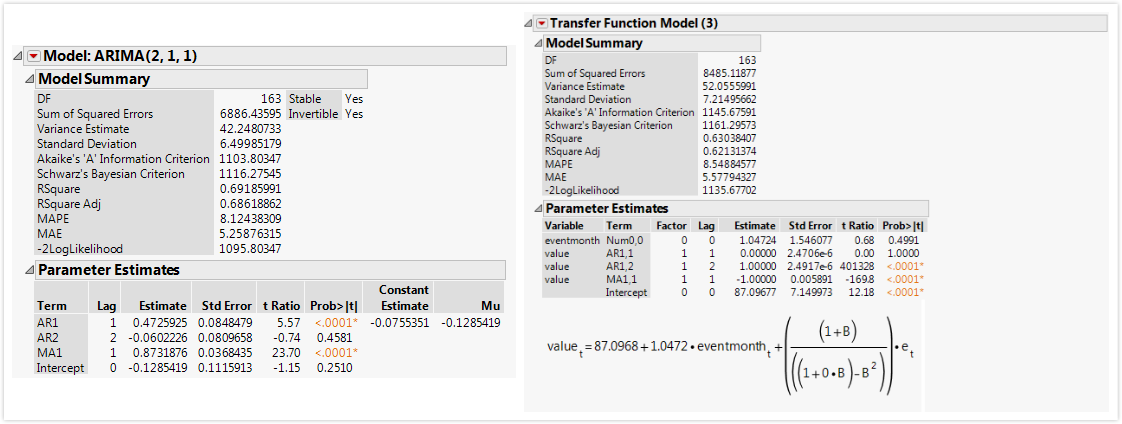}
\end{center}
\vspace{-.1in}
\caption[search engines]{Depression}
\label{stationmapb}
\end{figure}

\begin{figure}[!h]
\begin{center}
\leavevmode
\includegraphics[height=1in]{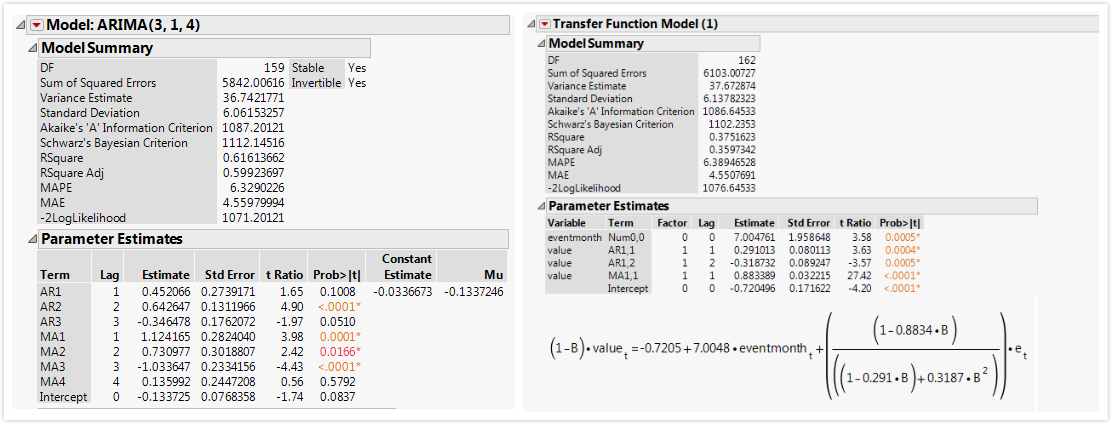}
\end{center}
\vspace{-.1in}
\caption[search engines]{Diabetes}
\label{stationmapb}
\end{figure}

\begin{figure}[!h]
\begin{center}
\leavevmode
\includegraphics[height=1in]{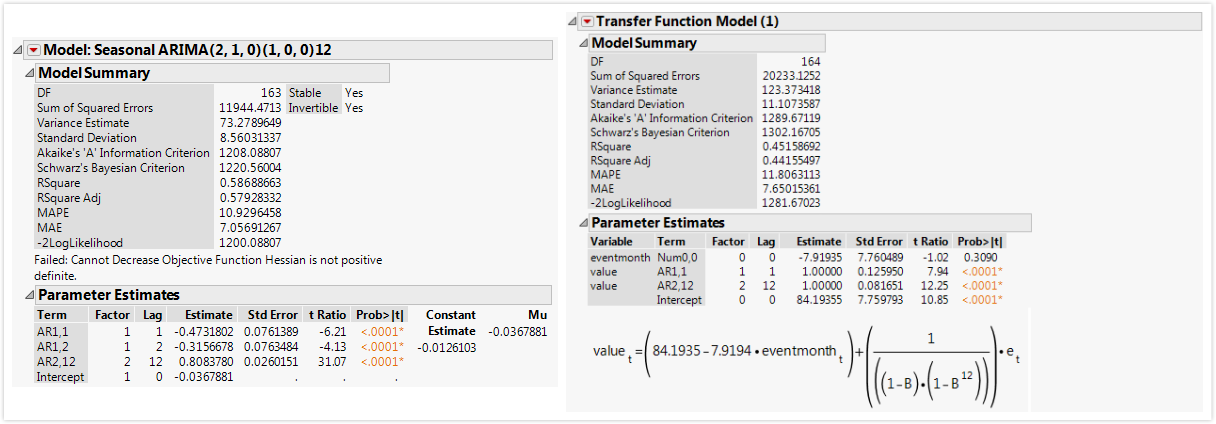}
\end{center}
\vspace{-.1in}
\caption[search engines]{Down Syndrome}
\label{stationmapb}
\end{figure}

\begin{figure}[!h]
\begin{center}
\leavevmode
\includegraphics[height=1in]{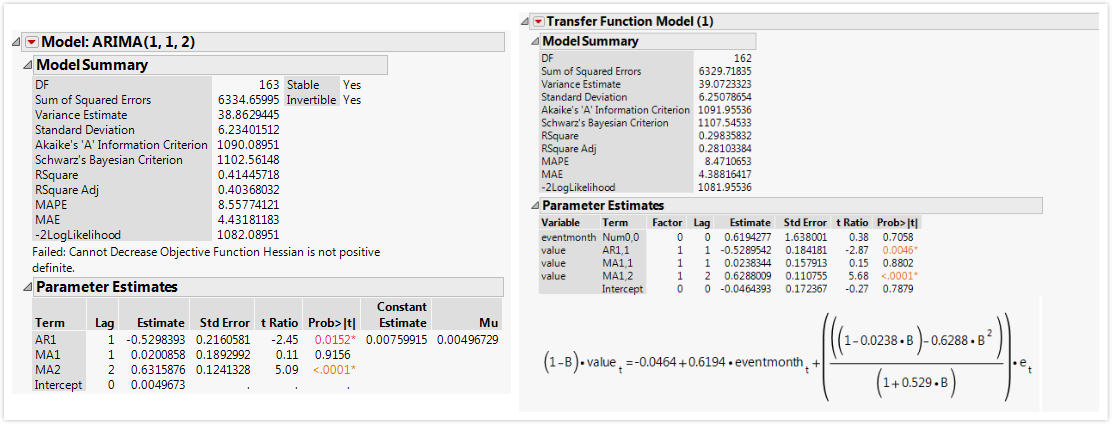}
\end{center}
\vspace{-.1in}
\caption[search engines]{Endometriosis}
\label{stationmapb}
\end{figure}

\begin{figure}[!h]
\begin{center}
\leavevmode
\includegraphics[height=1in]{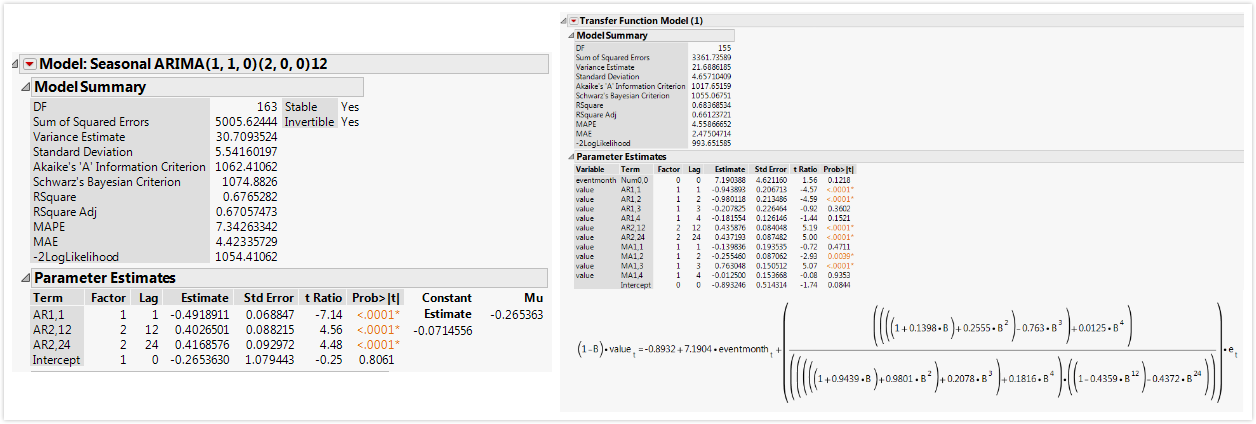}
\end{center}
\vspace{-.1in}
\caption[search engines]{Epilepsy}
\label{stationmapb}
\end{figure}

\begin{figure}[!h]
\begin{center}
\leavevmode
\includegraphics[height=1in]{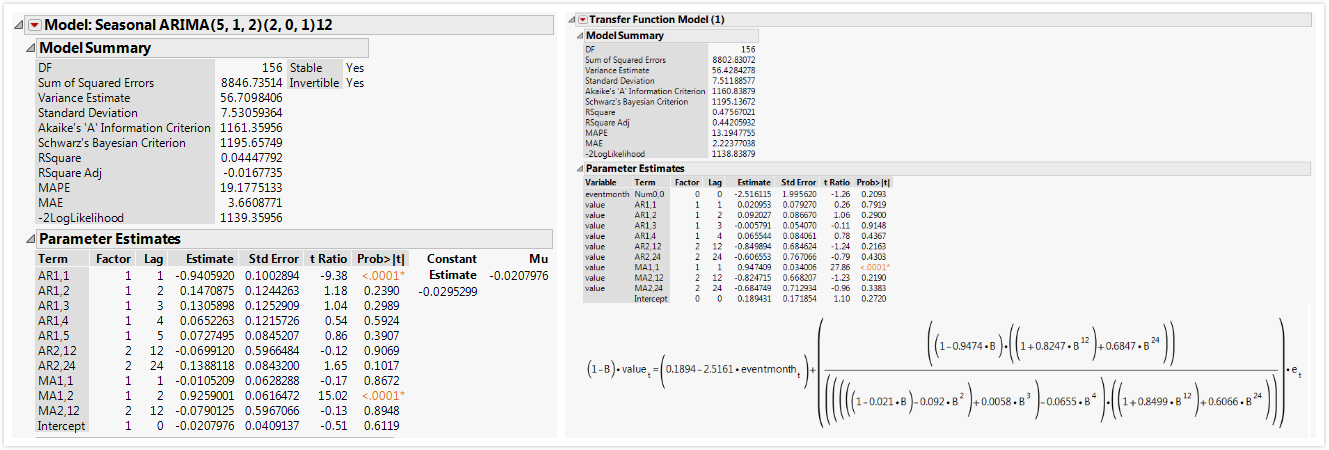}
\end{center}
\vspace{-.1in}
\caption[search engines]{Eye Injury}
\label{stationmapb}
\end{figure}

\begin{figure}[!h]
\begin{center}
\leavevmode
\includegraphics[height=1in]{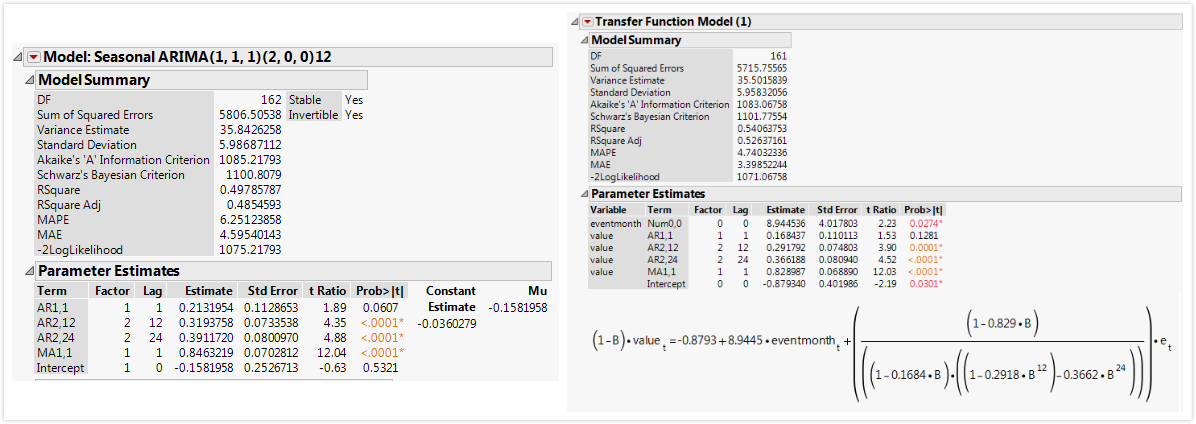}
\end{center}
\vspace{-.1in}
\caption[search engines]{Glaucoma}
\label{stationmapb}
\end{figure}

\begin{figure}[!h]
\begin{center}
\leavevmode
\includegraphics[height=1in]{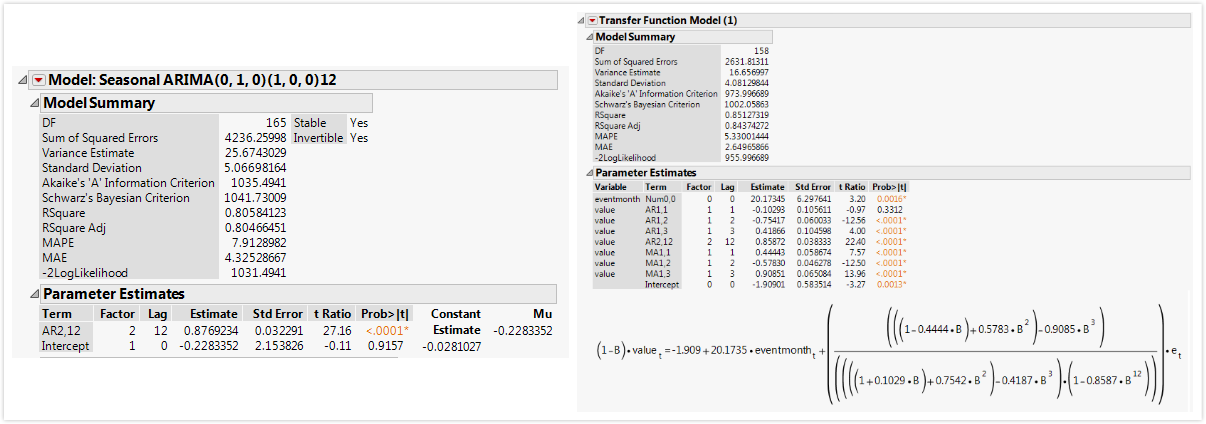}
\end{center}
\vspace{-.1in}
\caption[search engines]{Heart Disease}
\label{stationmapb}
\end{figure}

\begin{figure}[!h]
\begin{center}
\leavevmode
\includegraphics[height=1in]{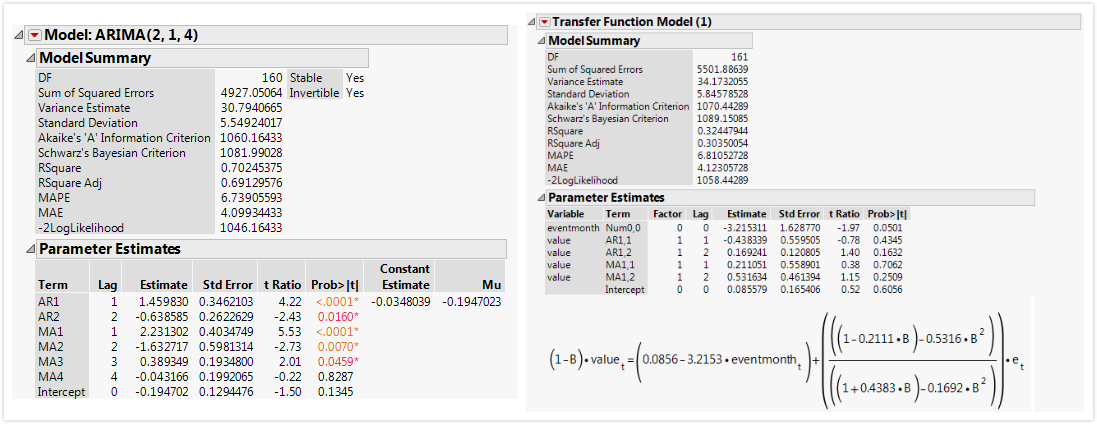}
\end{center}
\vspace{-.1in}
\caption[search engines]{Hepatitis}
\label{stationmapb}
\end{figure}

\begin{figure}[!h]
\begin{center}
\leavevmode
\includegraphics[height=1in]{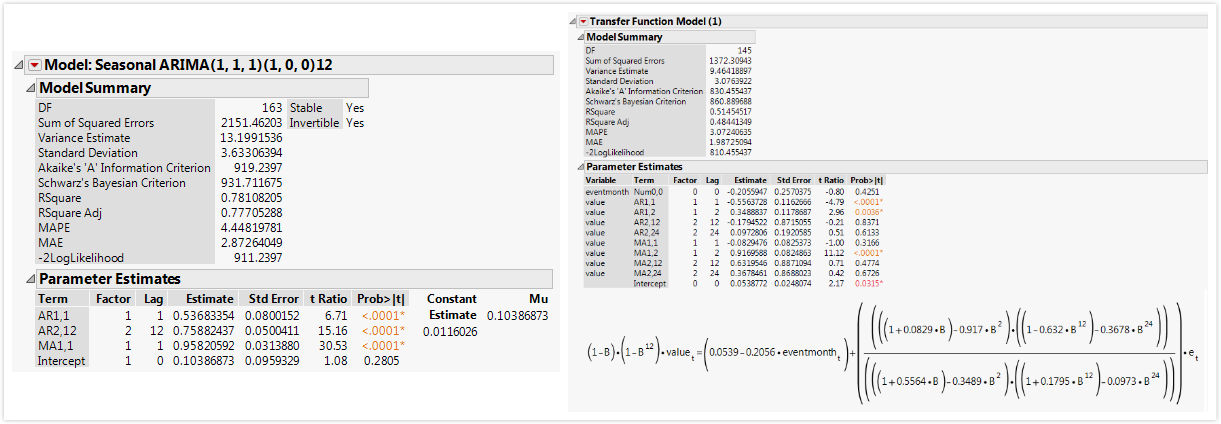}
\end{center}
\vspace{-.1in}
\caption[search engines]{High Blood Pressure}
\label{stationmapb}
\end{figure}

\begin{figure}[!h]
\begin{center}
\leavevmode
\includegraphics[height=1in]{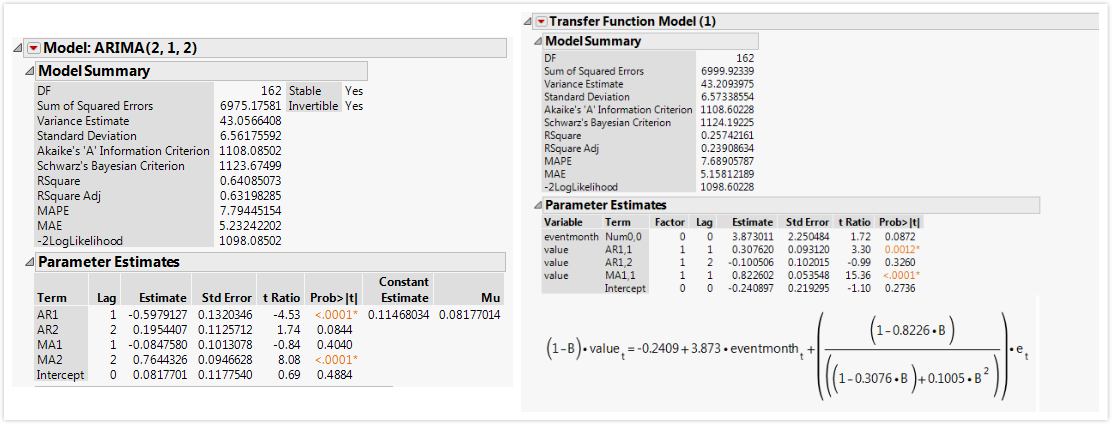}
\end{center}
\vspace{-.1in}
\caption[search engines]{Ibs}
\label{stationmapb}
\end{figure}

\begin{figure}[!h]
\begin{center}
\leavevmode
\includegraphics[height=1in]{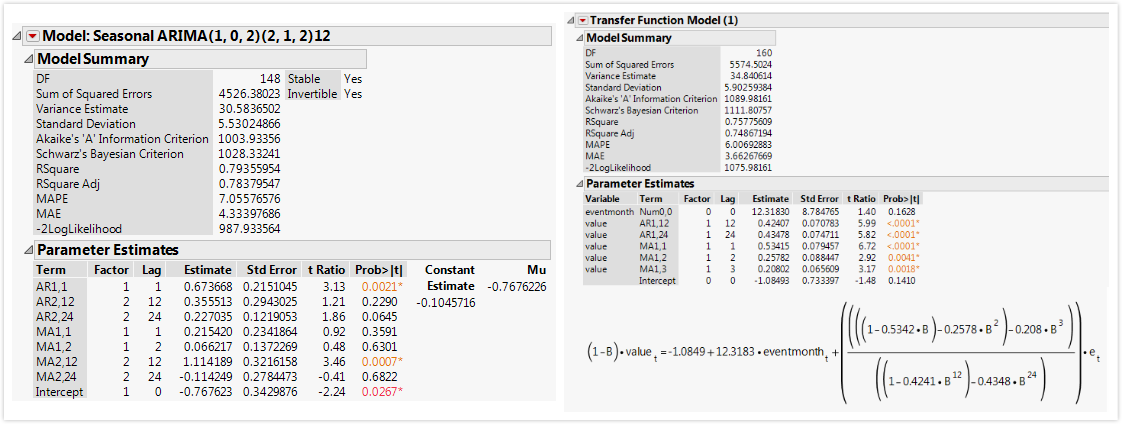}
\end{center}
\vspace{-.1in}
\caption[search engines]{Immunization}
\label{stationmapb}
\end{figure}

\begin{figure}[!h]
\begin{center}
\leavevmode
\includegraphics[height=1in]{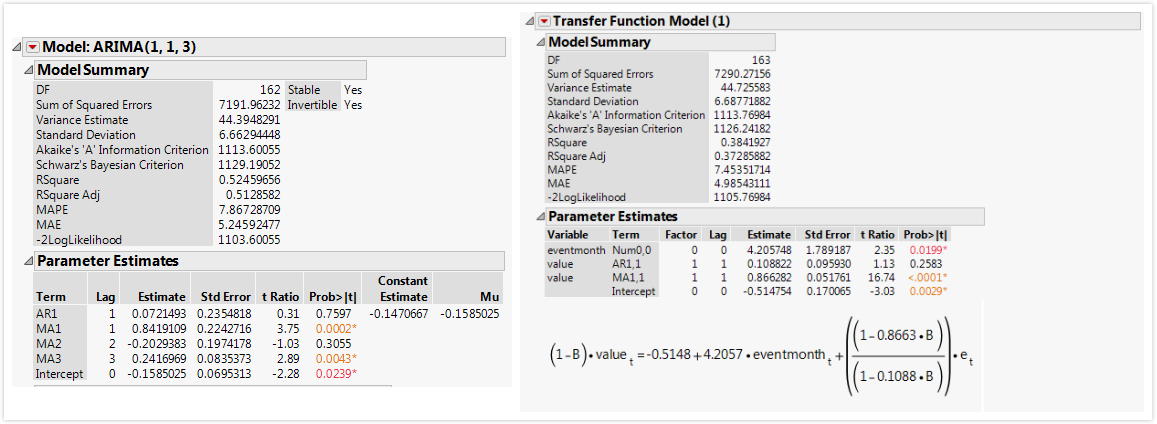}
\end{center}
\vspace{-.1in}
\caption[search engines]{Leukemia}
\label{stationmapb}
\end{figure}

\begin{figure}[!h]
\begin{center}
\leavevmode
\includegraphics[height=1in]{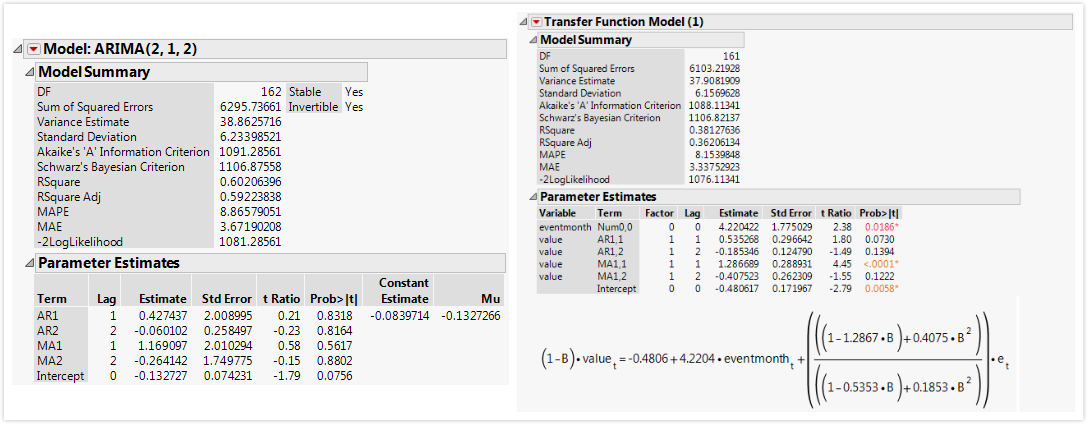}
\end{center}
\vspace{-.1in}
\caption[search engines]{Lung Cancer}
\label{stationmapb}
\end{figure}

\begin{figure}[!h]
\begin{center}
\leavevmode
\includegraphics[height=1in]{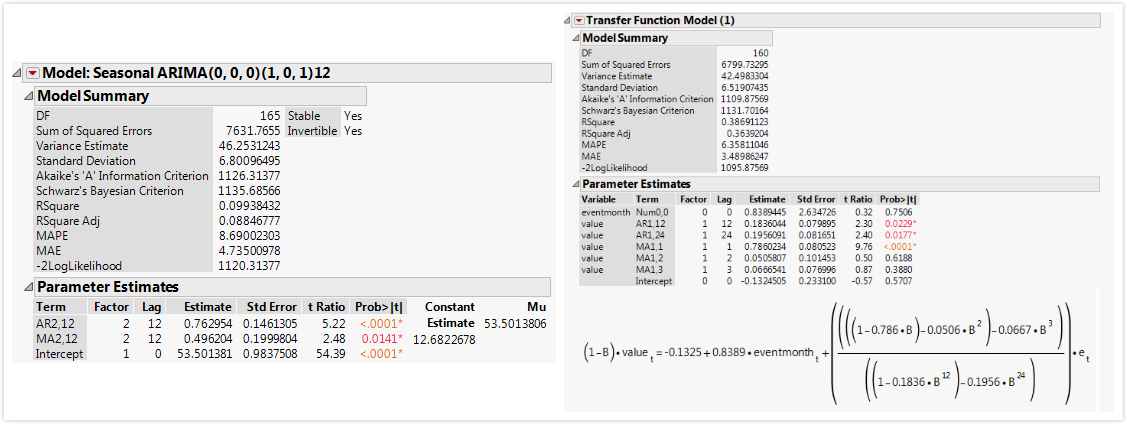}
\end{center}
\vspace{-.1in}
\caption[search engines]{Lupus}
\label{stationmapb}
\end{figure}

\begin{figure}[!h]
\begin{center}
\leavevmode
\includegraphics[height=1in]{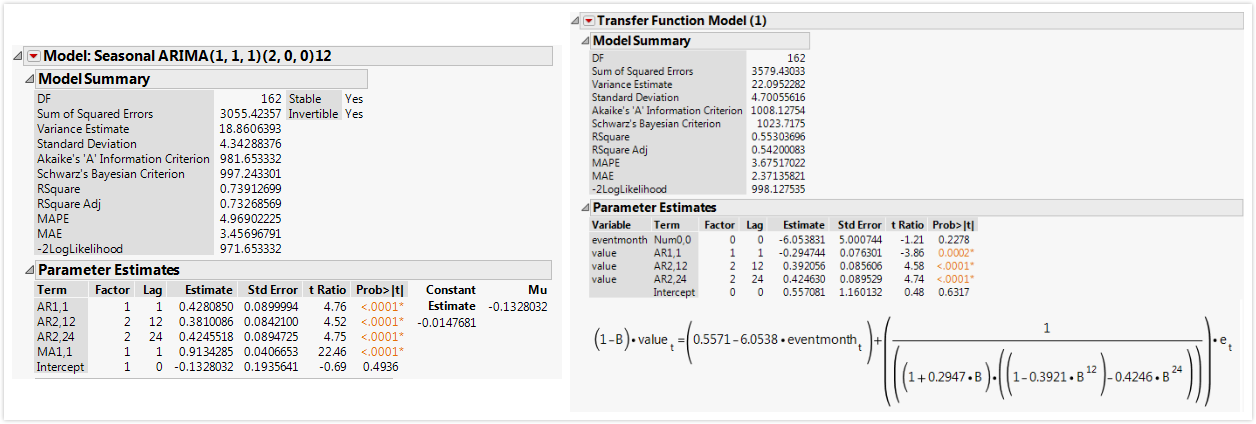}
\end{center}
\vspace{-.1in}
\caption[search engines]{Menopause}
\label{stationmapb}
\end{figure}

\begin{figure}[!h]
\begin{center}
\leavevmode
\includegraphics[height=1in]{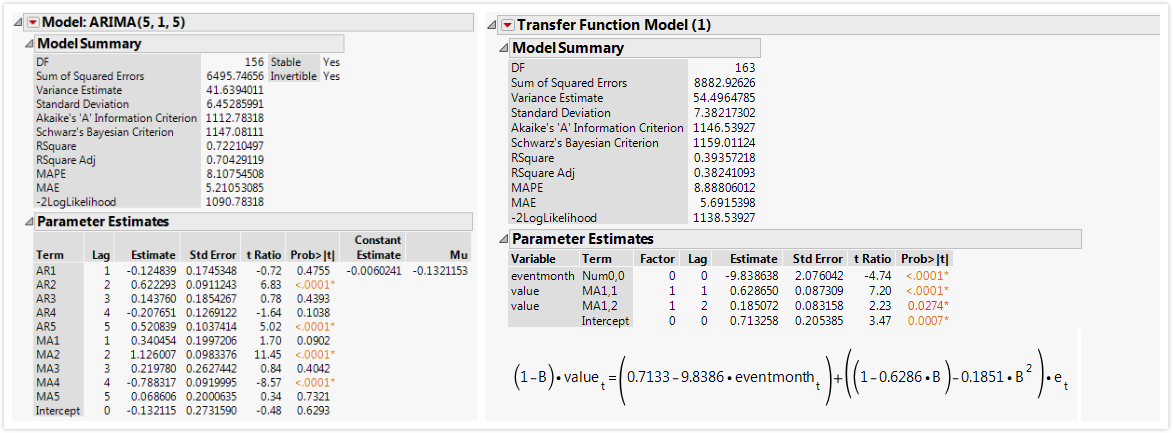}
\end{center}
\vspace{-.1in}
\caption[search engines]{Mental Health}
\label{stationmapb}
\end{figure}

\begin{figure}[!h]
\begin{center}
\leavevmode
\includegraphics[height=1in]{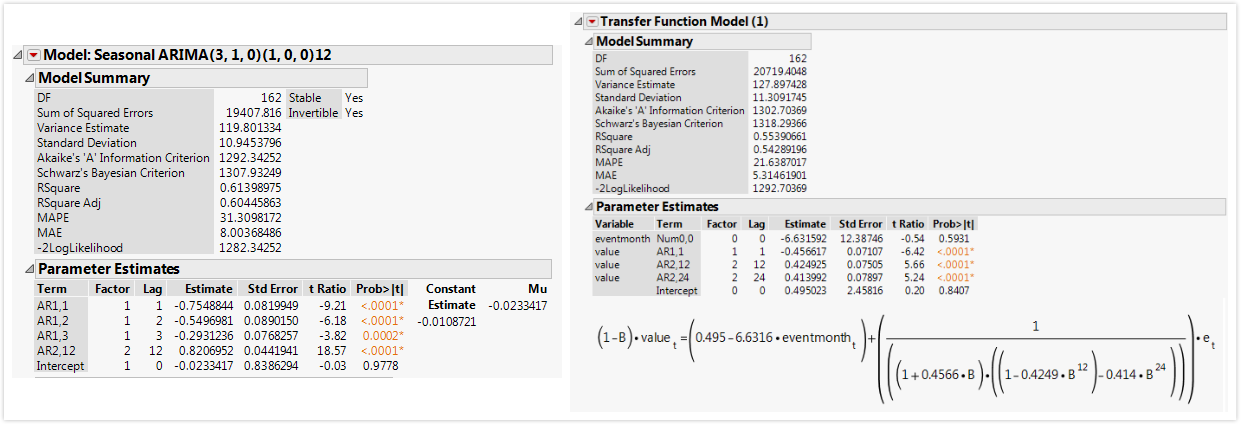}
\end{center}
\vspace{-.1in}
\caption[search engines]{National Nutrition}
\label{stationmapb}
\end{figure}

\begin{figure}[!h]
\begin{center}
\leavevmode
\includegraphics[height=1in]{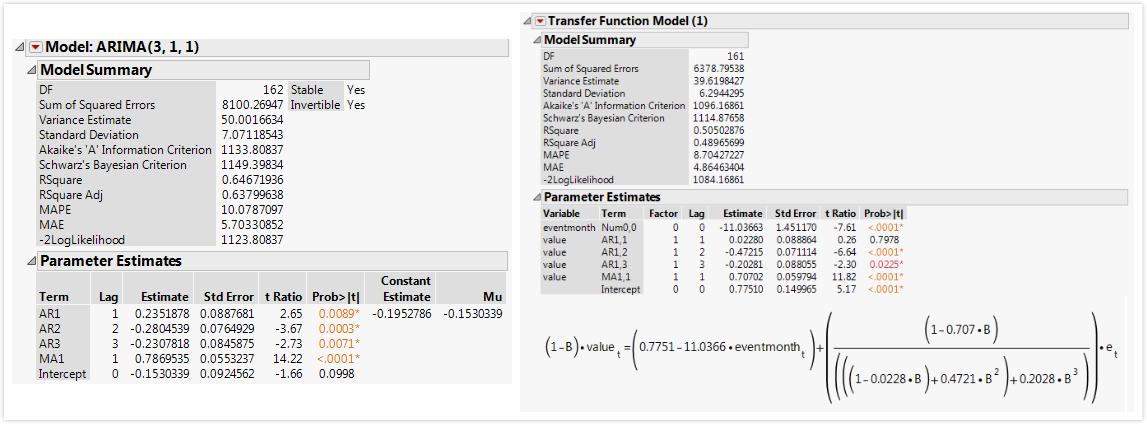}
\end{center}
\vspace{-.1in}
\caption[search engines]{Osteoporosis}
\label{stationmapb}
\end{figure}

\begin{figure}[!h]
\begin{center}
\leavevmode
\includegraphics[height=1in]{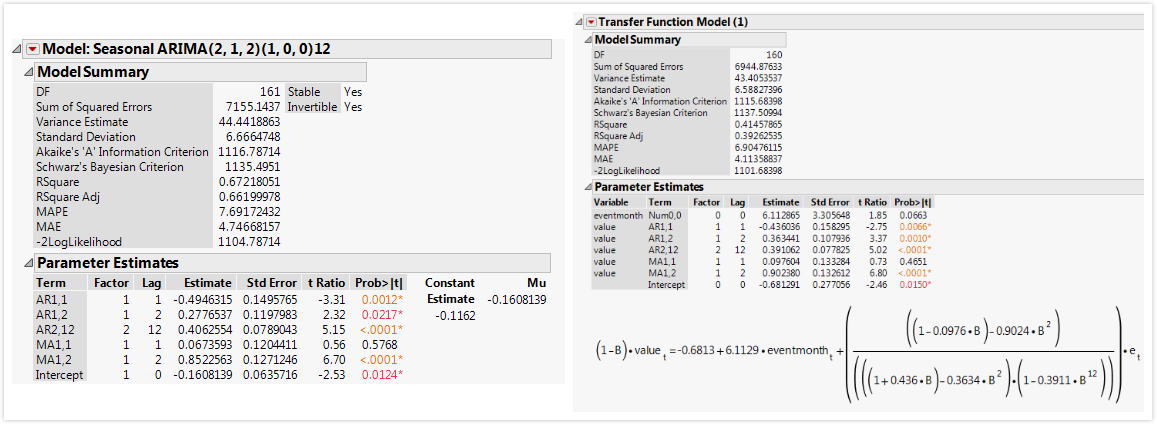}
\end{center}
\vspace{-.1in}
\caption[search engines]{Ovarian Cancer}
\label{stationmapb}
\end{figure}

\begin{figure}[!h]
\begin{center}
\leavevmode
\includegraphics[height=1in]{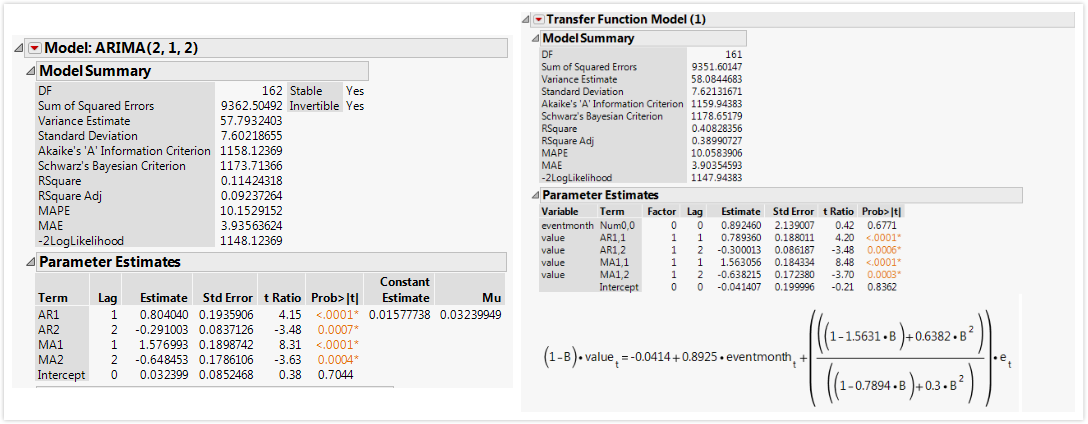}
\end{center}
\vspace{-.1in}
\caption[search engines]{Pancreatic Cancer}
\label{stationmapb}
\end{figure}

\begin{figure}[!h]
\begin{center}
\leavevmode
\includegraphics[height=1in]{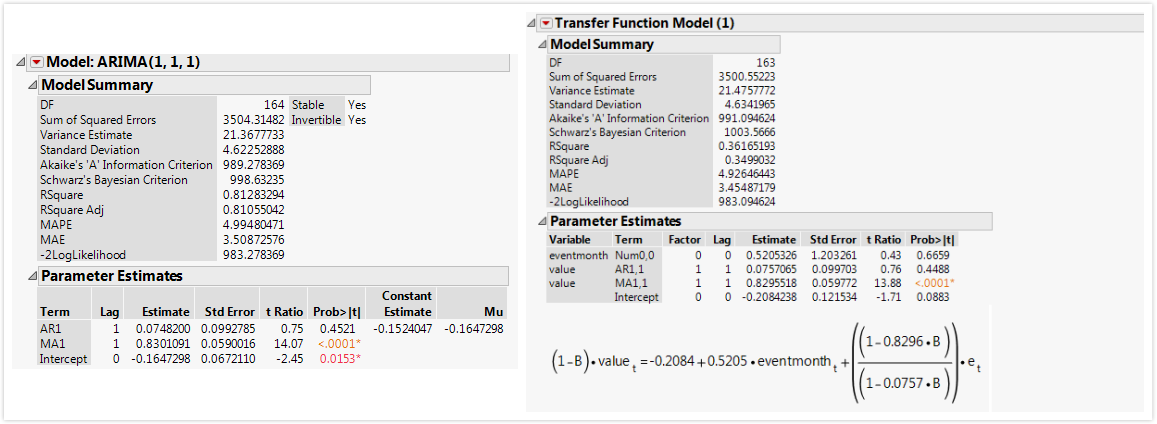}
\end{center}
\vspace{-.1in}
\caption[search engines]{Prostate}
\label{stationmapb}
\end{figure}

\begin{figure}[!h]
\begin{center}
\leavevmode
\includegraphics[height=1in]{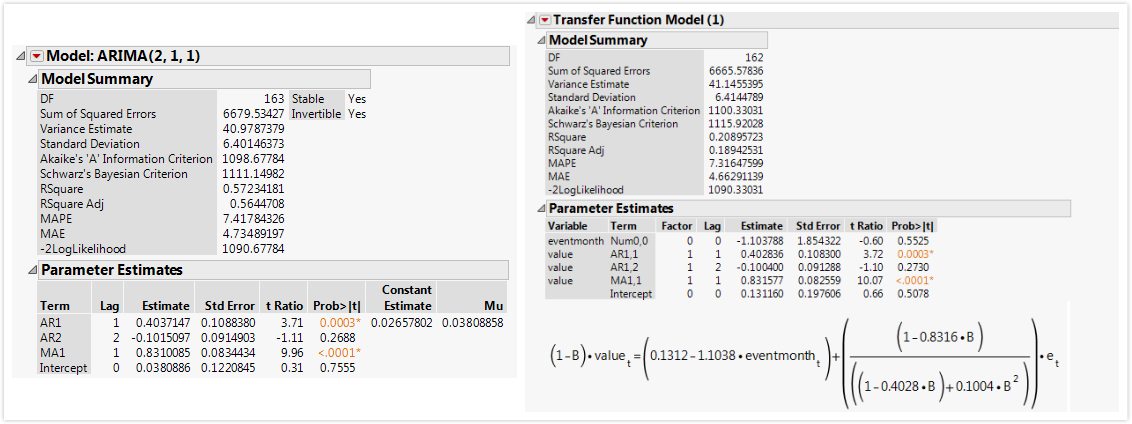}
\end{center}
\vspace{-.1in}
\caption[search engines]{Psoriasis}
\label{stationmapb}
\end{figure}

\begin{figure}[!h]
\begin{center}
\leavevmode
\includegraphics[height=1in]{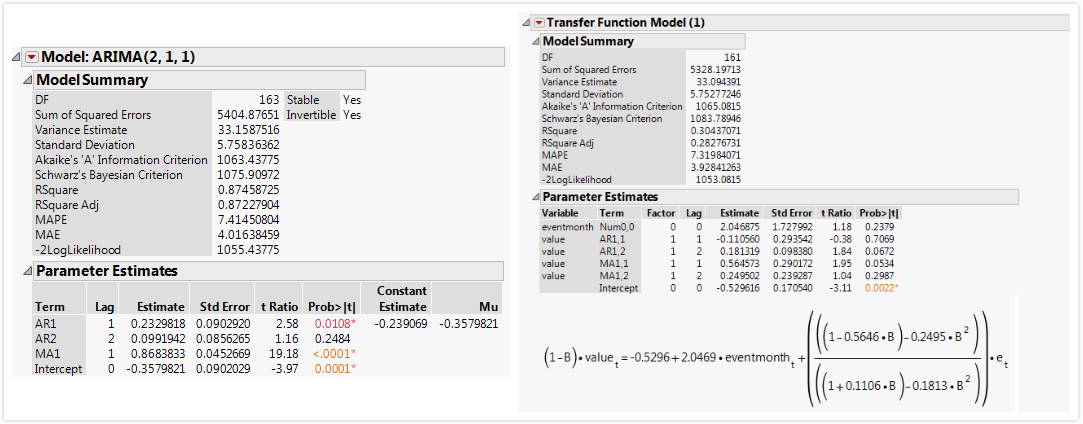}
\end{center}
\vspace{-.1in}
\caption[search engines]{Sclerosis}
\label{stationmapb}
\end{figure}

\begin{figure}[!h]
\begin{center}
\leavevmode
\includegraphics[height=1in]{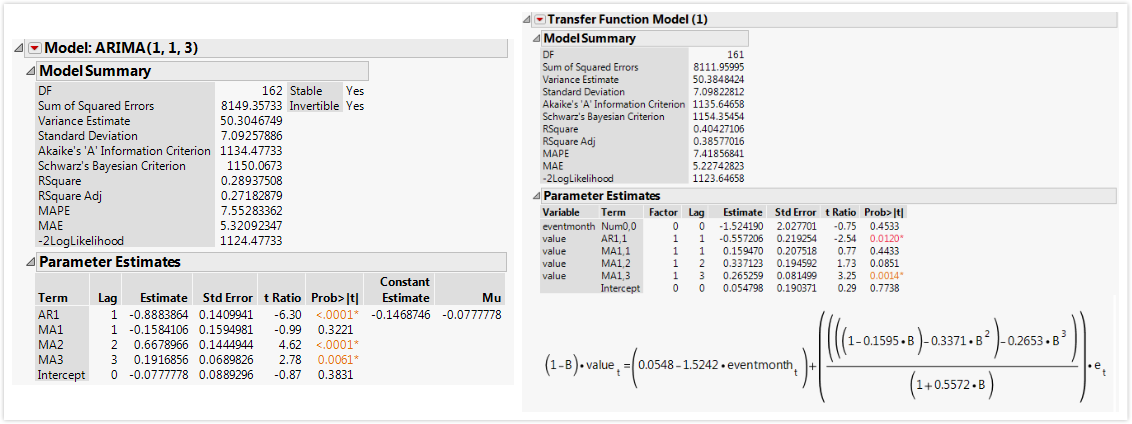}
\end{center}
\vspace{-.1in}
\caption[search engines]{Scoliosis}
\label{stationmapb}
\end{figure}

\begin{figure}[!h]
\begin{center}
\leavevmode
\includegraphics[height=1in]{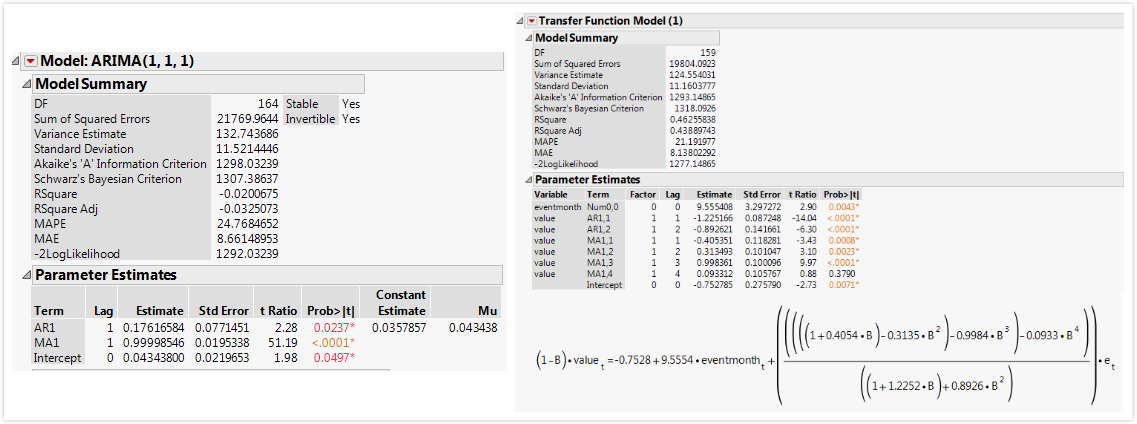}
\end{center}
\vspace{-.1in}
\caption[search engines]{Sids}
\label{stationmapb}
\end{figure}

\begin{figure}[!h]
\begin{center}
\leavevmode
\includegraphics[height=1in]{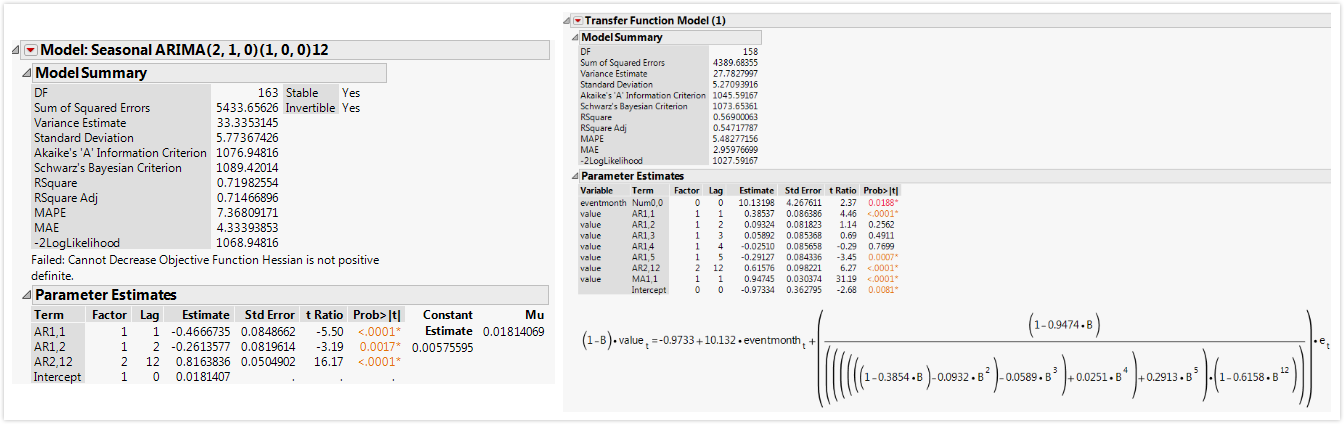}
\end{center}
\vspace{-.1in}
\caption[search engines]{Skin Cancer}
\label{stationmapb}
\end{figure}

\begin{figure}[!h]
\begin{center}
\leavevmode
\includegraphics[height=1in]{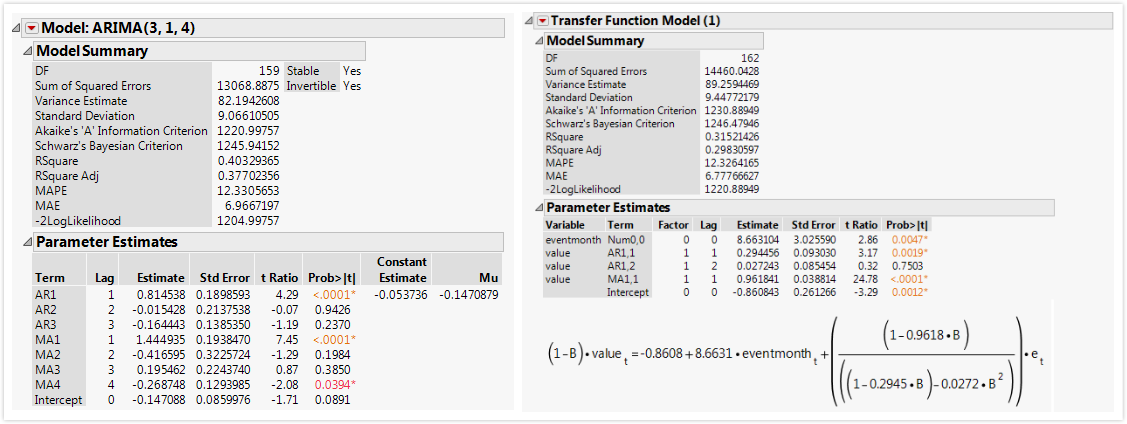}
\end{center}
\vspace{-.1in}
\caption[search engines]{Spina Bifida}
\label{stationmapb}
\end{figure}

\begin{figure}[!h]
\begin{center}
\leavevmode
\includegraphics[height=1in]{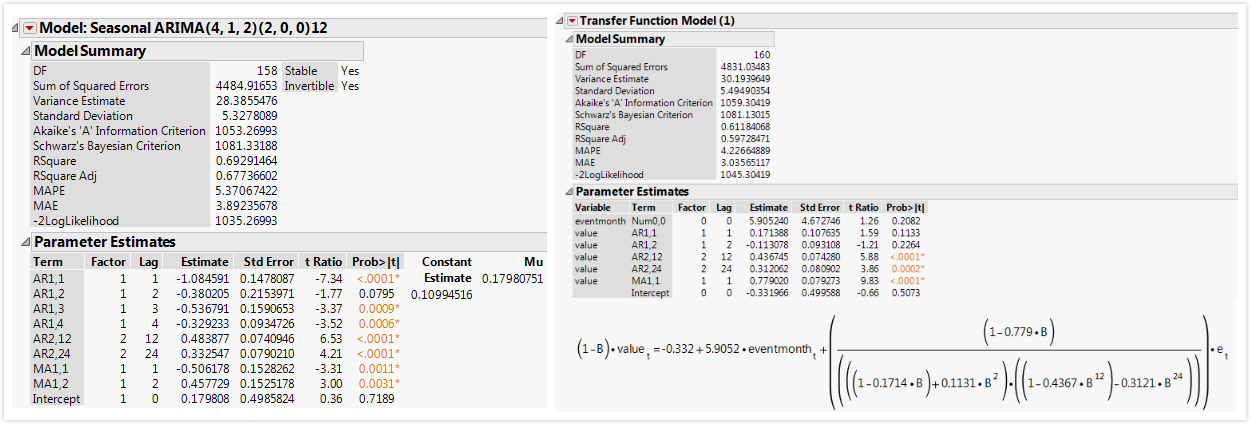}
\end{center}
\vspace{-.1in}
\caption[search engines]{Stroke}
\label{stationmapb}
\end{figure}

\begin{figure}[!h]
\begin{center}
\leavevmode
\includegraphics[height=1in]{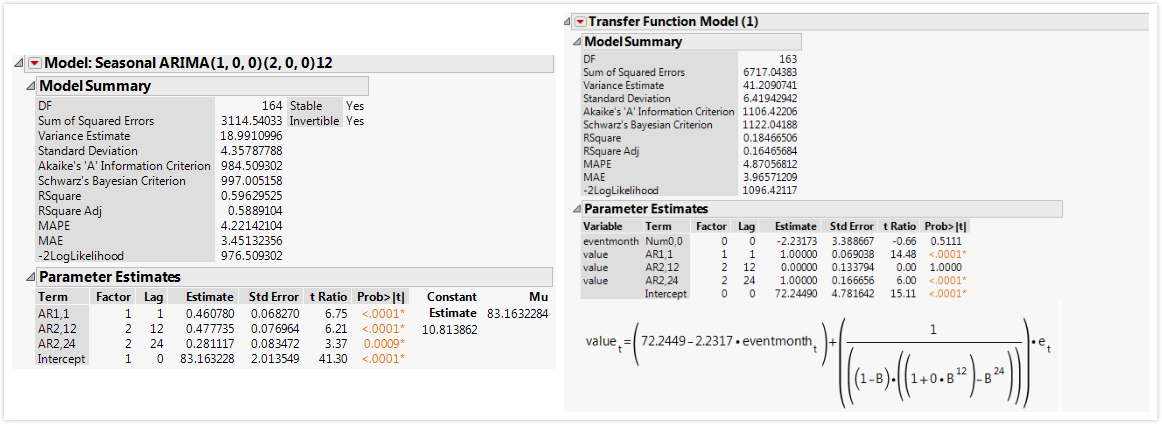}
\end{center}
\vspace{-.1in}
\caption[search engines]{Thyroid}
\label{stationmapb}
\end{figure}

\end{document}